\documentclass{ws-ijmpd}
\usepackage[super,compress]{cite}
\begin{document}

\markboth{C. Burigana et al.}
{Cosmology and fundamental physics with microwave surveys}

%%%%%%%%%%%%%%%%%%%%% Publisher's Area please ignore %%%%%%%%%%%%%%%
%
\catchline{}{}{}{}{}
%
%%%%%%%%%%%%%%%%%%%%%%%%%%%%%%%%%%%%%%%%%%%%%%%%%%%%%%%%%%%%%%%%%%%%

\title{RECENT RESULTS AND PERSPECTIVES ON COSMOLOGY AND FUNDAMENTAL PHYSICS FROM MICROWAVE SURVEYS\footnote{Based on presentations at the Fourteenth Marcel Grossmann Meeting on General Relativity, Rome, July 2015.}}

\author{CARLO BURIGANA$^{1,2,3,a}$, ELIA STEFANO BATTISTELLI$^{4,b}$, MICOL BENETTI$^{5,c}$,\\ 
GIOVANNI CABASS$^{4,d}$, PAOLO DE BERNARDIS$^{4,e}$,\\ 
SPERELLO DI SEREGO ALIGHIERI$^{6,f}$, ELEONORA DI VALENTINO$^{7,g}$,\\
MARTINA GERBINO$^{8,9,4,h}$, ELENA GIUSARMA$^{4,i}$, ALESSANDRO GRUPPUSO$^{1,3,j}$,\\ 
MICHELE LIGUORI$^{10,11,k}$, SILVIA MASI$^{4,l}$, HANS ULRIK NORGAARD-NIELSEN$^{12,m}$,\\ 
PIERO ROSATI$^{2,n}$, LAURA SALVATI$^{4,o}$,\\
TIZIANA TROMBETTI$^{1,2,p}$, PATRICIO VIELVA$^{13,q}$}
\address{$^{1}$INAF--IASF Bologna, Via Piero Gobetti 101, I-40129 Bologna, Italy
\footnote{Istituto Nazionale di Astrofisica --
Istituto di Astrofisica Spaziale e Fisica Cosmica di Bologna, Via Piero Gobetti 101, I-40129 Bologna, Italy}\\
$^{2}$Dipartimento di Fisica e Scienze della Terra, Universit\`a degli Studi di Ferrara,\\
Via Giuseppe Saragat 1, I-44122 Ferrara, Italy\\
$^{3}$INFN, Sezione di Bologna, Via Irnerio 46, I-40126, Bologna, Italy\\
$^{4}$Dipartimento di Fisica e INFN, Universit\`a di Roma ``La Sapienza'', \\P.le Aldo Moro 2, 00185, Rome, Italy\\
$^{5}$Observat\'orio Nacional, 20921-400, Rio de Janeiro, RJ, Brazil\\
$^{6}$INAF--Osservatorio Astrofisico di Arcetri, Largo Enrico Fermi 5, I-50125 Firenze, Italy\\
$^{7}$Institut d'Astrophysique de Paris \\ (UMR7095: CNRS \& UPMC -- Sorbonne Universities), F-75014, Paris, France\\
$^{8}$The Oskar Klein Centre for Cosmoparticle Physics, Department of Physics,\\
 Stockholm University, AlbaNova, SE-106 91 Stockholm, Sweden\\
$^{9}$Nordita (Nordic Institute for Theoretical Physics), Roslagstullsbacken 23, SE-106 91 Stockholm, Sweden\\
$^{10}$Dipartimento di Fisica e Astronomia G. Galilei,  Universit\`a degli Studi di Padova,\\ 
Via Marzolo 8, 35131 Padova, Italy\\
$^{11}$INFN, Sezione di Padova, via Marzolo 8, I-35131 Padova, Italy\\
$^{12}$DTU Space, Elektrovej, DK - 2800 Kgs. Lyngby, Denmark\\
$^{13}$Instituto de F{\'\i}sica de Cantabria (CSIC-UC), Santander, 39005, Spain\\
$^{a}$burigana@iasfbo.inaf.it $-$ $^{b}$elia.battistelli@roma1.infn.it $-$ $^{c}$micolbenetti@on.br\\
$^{d}$giovanni.cabass@gmail.com $-$ $^{e}$paolo.debernardis@roma1.infn.it $-$ $^{f}$sperello@arcetri.astro.it\\
$^{g}$valentin@iap.fr $-$ $^{h}$martina.gerbino@uniroma1.it $-$ $^{i}$elena.giusarma@roma1.infn.it\\
$^{j}$gruppuso@iasfbo.inaf.it $-$ $^{k}$michele.liguori@pd.infn.it $-$ $^{l}$silvia.masi@roma1.infn.it\\
$^{m}$hunn@space.dtu.dk $-$ $^{n}$rosati@fe.infn.it $-$ $^{o}$laura.salvati@roma1.infn.it\\
$^{p}$trombetti@iasfbo.inaf.it $-$ $^{q}$vielva@ifca.unican.es
}

\maketitle

\begin{history}
\received{1 March 2016}
\accepted{2 March 2016}
%\revised{2 March 2016}
\end{history}

\begin{abstract}
Recent cosmic microwave background data in temperature and polarization 
have reached high precision in estimating all the parameters that describe the  
current so-called standard cosmological model. Recent results about the integrated Sachs-Wolfe effect from cosmic microwave background anisotropies, galaxy surveys,
and their cross-correlations are presented.
Looking at fine signatures in the cosmic microwave background, such as the lack of power at low multipoles, the primordial power spectrum and the bounds on non-Gaussianities, 
complemented by galaxy surveys, we discuss inflationary physics and the generation of primordial perturbations in the early Universe.
Three important topics in particle physics, 
the bounds on neutrinos masses and parameters, on thermal axion mass and on the neutron lifetime derived from cosmological data
are reviewed, with attention to the comparison with laboratory experiment results.
Recent results from cosmic polarization rotation analyses aimed at testing the Einstein equivalence principle are presented.
Finally, we discuss the perspectives of next radio facilities for the improvement of
the analysis of future cosmic microwave background spectral distortion experiments.
\end{abstract}

% for keywords & PACS numbers of AIP see:
% https://www.aip.org/publishing/pacs/pacs-reg90#95


\ccode{PACS numbers: 98.80.-k; 98.70.Vc; 95.85.Bh; 98.80.Bp; 98.80.Cq; 98.80.Es; 98.65.Dx; 95.35.+d; 95.36.+x; 95.30.Cq.}

%\tableofcontents

\def\lsim{\,\lower2truept\hbox{${< \atop\hbox{\raise4truept\hbox{$\sim$}}}$}\,}
\def\gsim{\,\lower2truept\hbox{${> \atop\hbox{\raise4truept\hbox{$\sim$}}}$}\,}

\section{Introduction}

Latest measurements of cosmic microwave background (CMB) anisotropies in temperature and polarization 
from {\it Planck} satellite,\footnote{{\it Planck} is a project of the European Space Agency - ESA - with instruments provided by
two scientific Consortia funded by ESA member states (in particular the lead countries: France and Italy) with
contributions from NASA (USA), and telescope reflectors provided in a collaboration between ESA and a scientific
Consortium led and funded by Denmark.}\cite{Planck:2015xua}
complemented at smaller scales by recent ground-based experiments
(see e.g. Refs. [\refcite{Das:2013zf,Ade14c,2015ApJ...799..177G,2015PhRvL.114j1301B}]) 
and combined with other cosmological information coming from e.g. type-Ia supernovae, 
galaxy and galaxy cluster surveys, have reached high precision in estimating all the parameters that describe the 
current so-called standard cosmological model. Far from representing a fully, physically exhaustive interpretation of the Universe properties,
the cosmological constant plus cold dark matter ($\Lambda$CDM) model phenomenologically describes 
%very
reasonably 
well existing data with a simple set of six parameters (see e.g. the lectures by M. Bersanelli and J.-L. Puget on {\it Planck}
results\footnote{This paper is based largely on the products available at the ESA {\it Planck} Legacy Archive
and publicly available publications by ESA and the {\it Planck}
Collaboration, for what concerns the related aspects. Any material
presented here that is not already described in {\it Planck}
Collaboration papers represents the views of the authors and not
necessarily those of the {\it Planck} Collaboration.} in this Meeting).
The integrated Sachs-Wolfe effect, discussed here in Sect. \ref{ISWsection}, represents a remarkable example of the success of current cosmology, 
since a such intrinsically weak predicted signal is clearly recognized in two classical cosmological probes, like CMB anisotropies and galaxy surveys,
and in their cross-correlations.
Looking at fine signatures in the CMB it is possible to derive more hints on early Universe and inflationary physics as well as to 
carry out a sort of laboratory tests to constrain particle and fundamental physics. 
In Sect. \ref{aba:sec1} the ``lack of power'' in the large scale pattern (i.e. low multipole region) of CMB anisotropy angular power spectrum (APS) 
is investigated to link the inflationary phase to the string theory while CMB data and galaxy surveys are jointly analyzed in Sect. \ref{features} to  
constrain inflationary models predicting localized `features' in the primordial power spectrum (PPS). 
Going beyond power spectrum (PS) analyses, the study of primordial non-Gaussianity (PNG), discussed in Sect. \ref{nongauss}, 
allows to test mechanisms for the generation of primordial perturbations in the early Universe. 
Sects. \ref{neutrinossec} and \ref{axionssec} discuss two important topics in dark matter studies, respectively 
the bounds on neutrinos masses and parameters and on thermal axion mass from cosmological data
while Sect. \ref{neutronsec} summarizes the state of the art on the neutron lifetime, $\tau_n$, a fundamental quantity in nuclear physics.
Attention is given to the comparison with laboratory experiment results.
Sect. \ref{cprsec} is devoted to the test of the Einstein equivalence principle (EEP), at the basis of general relativity (GR), through the analysis of
the cosmic polarization rotation (CPR) and to the comparison of results from astronomical and CMB based analyses.
Finally, Sect. \ref{sec:cmb_spect} discusses the main cosmological and fundamental physics information contained in the CMB spectral distortions in the light 
of the contribution expected from the Square Kilometre Array (SKA).

\def\planck{\textit{Planck}}

\section{Integrated Sachs-Wolfe effect}
\label{ISWsection}

The late integrated Sachs-Wolfe (ISW) effect\cite{Sachs1967, Rees1968, Martinez1990b} is a secondary anisotropy in the cosmic microwave background (CMB), which is caused by the interaction of CMB photons with the time-dependent gravitational potential of the evolving cosmic large-scale structure (LSS).
The ISW effect can be generated under several scenarios affecting the late evolution of the structures: a cosmological constant, dark energy (DE),\cite{Crittenden1996} modified gravity,\cite{Hu2002a} 
or spatial curvature.\cite{Kamionkowski1996}

\renewcommand{\arraystretch}{1.5}

%%%%%%%%%%%%%%%%%%%%%%%%%%%%%%%%%%%%%%%%%%%%%%%%%%%%%%%%%%%%%%%%%%%%%%%%%%%%%%%%%

%\subsection{The ISW effect} \label{sec:isw-theory}

The early ISW is generated after recombination (since the energy density of relativistic matter is still considerable at that time):
it adds in phase with the Sachs-Wolfe primary anisotropy, increasing the height of the first acoustic peaks. 
Besides, the effect %of $\Theta_\ell^{e\text{ISW}}(k)$ 
on the APS, $C_\ell$ (being $\ell$ the multipole of the spherical harmonic expansion), is suppressed by the factor $\rho^2_\textup{rad}(\eta_\textup{rec})/\rho^2_\textup{m}(\eta_\textup{rec})$: increasing the radiation energy density
with respect to that of matter near recombination %(by taking the effective number of relativistic species $N_\textrm{eff} > 3.046$)
will %delay the advent of matter domination and 
give a larger early ISW effect. The late ISW effect is active at more recent times: %, when dark energy starts to play a role and the gravitational potentials are decreasing.
focusing on scales corresponding to galaxy clusters, %where gravitational perturbations start growing, 
the CMB photons get redshifted by the time-dependent gravitational potentials. 
The %se 
potentials causing the late ISW also give rise to the weak lensing distortions: the interplay between these two effects results in a non-Gaussian correlation between small and large angular scales, which is encoded in the lensing-induced bispectrum.

\def\planck{\textit{Planck}}

The optimal detection\cite{Crittenden1996} of the ISW effect is made by the cross-correlation of the CMB temperature anisotropies with tracers of the gravitational potential, like, for instance, galaxy catalogues.
The first detection\cite{Boughn2004} was made using 
Wilkinson Microwave Anisotropy Probe (WMAP) data and radio and X-ray galaxy catalogues.
%After this work, several analyses followed providing additional evidence of the ISW detection using different surveys and correlation techniques. 
The ISW signal is very weak (an ideal LSS tracer could provide a detection of up to $\approx 8\,\sigma$), and, therefore, its capability to constrain cosmological parameters, is very limited. 
Nevertheless, using the ISW signal alone it is possible to constrain some cosmological parameters, by fixing the remaining ones to their standard value
%, helping to confirm the standard \lcdm\ model. Examples of these determinations are, for instance, 
(see e.g. the estimation of the DE density parameter\cite{planck2015-isw} $\Omega_\Lambda \approx 0.67$, with an error of about 20\%; the compatibility of the DE equation of state parameter with the expected value for a
$\Lambda$CDM scenario;\cite{Vielva2006} or the setting of upper limits on spatial flatness of a few per cent\cite{Li2010}).

We will focus here on the main results of the ISW effect derived by \planck\ (see Refs. [\refcite{planck2013-isw}] and [\refcite{planck2015-isw}] for a complete description).

%\subsection{CMB-LSS cross-correlation}

\subsection{The ISW probed through the CMB-LSS cross-correlation}

The four CMB maps\cite{planck2015-cmb} produced by  \planck\ 
({\tt COMMANDER}, {\tt NILC}, {\tt SEVEM}, {\tt SMICA})
%(\cruler, \nilc, \sevem, and \smica) 
have been cross-correlated with several tracers of the LSS. In the first release, 
the NRAO VLA Sky Survey (NVSS) radio-galaxy catalogue, the photometric luminous galaxy (SDSS-CMASS/LOWZ), and the photometrically-selected galaxies (SDSS-MphG) from the 
%\textit{Sloan Digital Sky Survey} 
Sloan Digital Sky Survey 
(SDSS) were considered. Two additional catalogues from the 
%\textit{Wide-Field Infrared Survey Explorer} 
Wide-Field Infrared Survey Explorer
(WISE) were added to the analysis of the second release: one based on star-forming galaxies (WISE-GAL), and another one based on active galactic nuclei (WISE-AGN).
Considering the full cross-correlations of the CMB with all the LSS tracers, the latest results provided a total ISW detection
of around $3\,\sigma$, as expected for the standard $\Lambda$CDM model. The NVSS catalogue already provides by itself a similar detection level.

The most novel result provided by \planck\ 
%in relation to these cross-correlations, 
was its capability to provide a detection of the ISW 
%by itself, 
without relying on external tracers of the LSS,
%. This was possible 
thanks 
%due 
to
its
%the \planck\ 
%capabilities to provide a 
reliable estimation of the gravitational potential through the lensing suffered by the CMB photons.\cite{planck2015-lens}
The cross-correlation of this map with the CMB one, or, equivalently, the specific shape of the ISW-lensing bispectrum, reported a detection of the ISW at around $\approx 3\,\sigma$. 
%This is similar to what is obtained by cross-correlating the CMB with the external tracers. In fact, when 
When all the LSS tracers are combined, the total ISW detection is $\approx 4\,\sigma$, also in good agreement with the $\Lambda$CDM model.

%\subsection{Recovery of the ISW map}

\begin{figure}
\begin{center}
\includegraphics[width=2.2in]{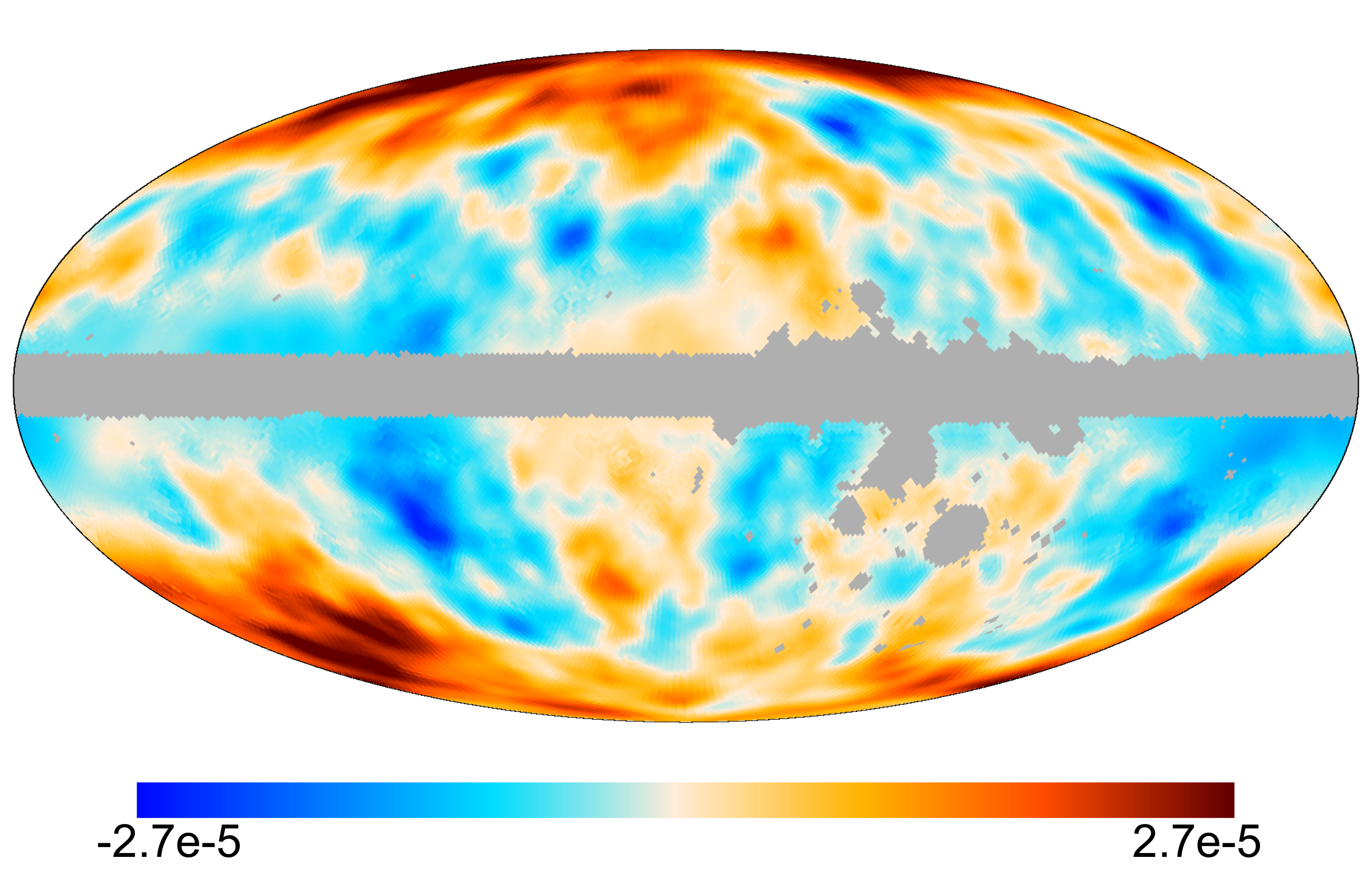}
\includegraphics[width=2.2in]{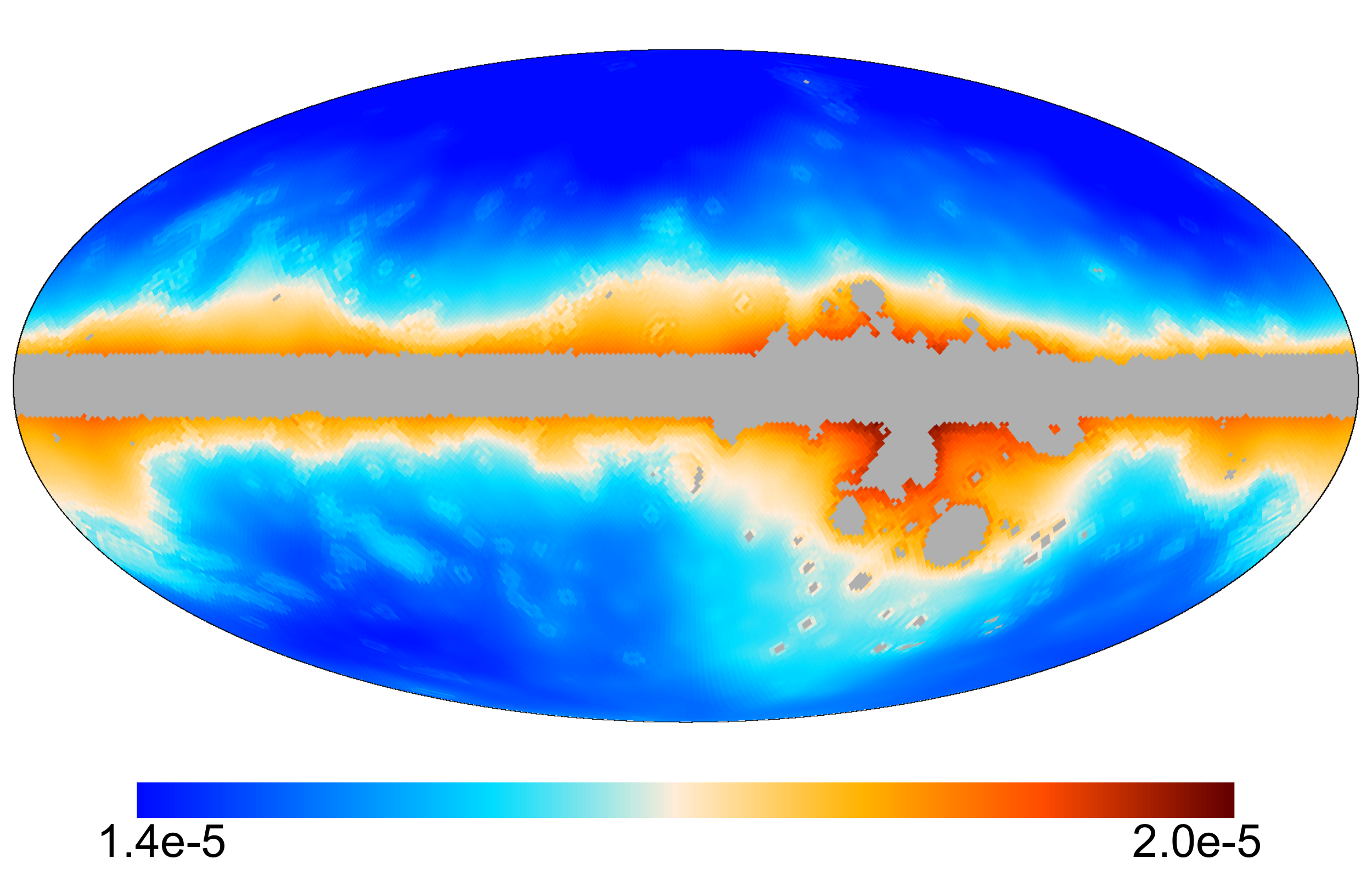}
\end{center}
\caption{Map (thermodynamic temperature in K) of the recovered ISW anisotropies (left) and its corresponding estimated error per pixel (right), from the combination of the {\it Planck} {\tt SEVEM} CMB map and all the LSS tracers.
%(\nvss, \wagn, \wg, \lrg, \mg, and lensing). 
The error map structure is determined by the sky coverage of the different surveys.
The total signal-to-noise of the ISW map cannot exceed, obviously, a value of $\approx 4$. The signal-to-noise is higher near the Galactic poles, with values of $\approx 2$.}  
\label{fig:isw}
%\vskip -0.3cm
\end{figure}

Assuming the standard $\Lambda$CDM model, the statistical ISW captured in the CMB-LSS cross-correlation can be used 
to estimate a map of the ISW anisotropies caused by the gravitational potential traced by each of the LSS probes.\cite{Barreiro2008} 
%ISW maps associated to the LSS tracers mentioned in the previous section were provided by the \planck\ Collaboration. 
%As an example, in 
Fig.~\ref{fig:isw} %(left panel) 
shows the ISW %fluctuations 
fluctuation
maps
obtained from the full cross-correlation of the {\it Planck} {\tt SEVEM} CMB map with NVSS, WISE-AGN, WISE-GAL, SDSS-CMASS/LOWZ, SDSS-MphG, and the \planck\ lensing LSS tracers. 
%On the right panel, a map of the estimated error per pixel is also provided. 

%
%\begin{figure}[ht]

All 
these
%the 
results 
%summarized in this Section and in the previous one 
were obtained without using CMB polarization information (except for the estimation of the \planck\ lensing map). In principle, including polarization could increase the ISW detection\cite{Frommert2009} around a $15\%$, however, the current CMB polarization data from \planck\ is high-pass filtered at the largest angular scales ($\gtrsim 5^{\circ}$), which are the most important ones in this context.
%to detect and characterize the ISW effect. 
Including the large scale polarization is, perhaps, the most important remaining aspect within the context of the ISW study, at least, from the CMB side. On the other side, future galaxy surveys 
like {\it Euclid},\cite{Laureijs2011} 
%\textit{J-PAS}~\cite{Benitez2014} or \textit{LSST}~\cite{lsst2012}, 
J-PAS\cite{Benitez2014} or LSST,\cite{lsst2012}  
among others, will provide accurate galaxy catalogues, probing very large volumes, allowing to perform, for instance, a tomographic ISW detection.

%\subsection{Stacking of CMB fluctuations}

A complementary approach 
%to detect the ISW effect 
consists in stacking the CMB fluctuations in the position of known structures, such as voids and clusters,
%. This was 
as
done initially on WMAP data\cite{Granett2008}
%, 
using a catalogue (GR08) of super-structures from SDSS. 
%This study indicated 
An anomalous ISW signal, incompatible with the standard $\Lambda$CDM model
%. 
was found, 
%This has been 
confirmed in the \planck\ analyses, showing that the intensity of the detected signal ($\approx -11\,\mu$K for voids, and $\approx +8.5\, \mu$K for clusters) and the scale at which that signal is maximum 
($\approx 3.5^\circ$ for voids, and $\approx 4.5^\circ$ for clusters) are, indeed, unexpected.

At these scales, the current CMB \planck\ polarization map still retains certain signal, despite the high-pass filtering and, therefore, it can be used to test further the nature of this anomalous signal. The key point is that, if this signal is caused by the ISW effect, and, therefore, originated by a gravitational secondary CMB anisotropy, a negligible contribution of the CMB polarization is expected. In fact, no associated polarization is found in \planck\ data, although the diminishing of the signal caused by the high-pass filtering 
%does not allow one to derive 
limits any strong conclusion. 
Anyway, 
%It can only be said that 
the current polarization data are not in contradiction with assuming that the emission coming from these GR08 structures provides an anomalous ISW signal. Studying the stacked fluctuations of the \planck\ lensing map on the GR08 positions also supports this hypothesis. In fact, 
%it is shown that, 
at least for the voids, 
%there is 
a clear correlation between the lensing gravitational potential and the position of the super-structures is found.

This kind of studies
%, and the analyses summarized in the previous Sections, 
could be further extended once \planck\ provides its next and final release, which will include polarization information at all the angular scales.

\newcommand{\TT}{TT+\text{lowP}}
\newcommand{\tu}{\textup}

\renewcommand{\arraystretch}{1.5}

%%%%%%%%%%%%%%%%%%%%%%%%%%%%%%%%%%%%%%%%%%%%%%%%%%%%%%%%%%%%%%%%%%%%%%%%%%%%%%%%%

\subsection{Parametrization of early and late ISW and data analysis} \label{sec:parametrization+analysis}

%We consider a parametrization of 
The ISW amplitude can be parametrized in terms of $A_{e\text{ISW}}$ and $A_{l\text{ISW}}$, which rescale the contribution of the ISW to the temperature anisotropies. 
%We perform 
A Markov-chain Monte-Carlo (MCMC) analysis %: %making use of the publicly available code \texttt{cosmomc}. %\cite{Lewis:2013hha, Lewis:2002ah} 
was performed
%our 
with a baseline %is the 
standard $\Lambda$CDM model
%, and we impose 
and flat priors on the parameters.\cite{cabassetal2016} 
We also check the impact of a Gaussian prior %on 
$A_{l\text{ISW}} = 1.00\pm 0.25$, % (which will be denoted by the ``prior'' label in the following plots and tables). 
consistent with the 68\% confidence level (C.L.) bounds on the same parameter from the estimation of the ISW-lensing bispectrum, %induced on the Gaussian CMB anisotropies by %the deflection caused by
%the lensing effect is estimated by %. This bound has been derived by
which has been obtained by cross-correlating the \planck\ CMB maps with the \planck\ map of the lensing potential.
%We test the following 
Various datasets 
were tested: the high-$\ell$ \planck\ temperature and temperature+polarization APS in the range $30\leq\ell<2500$ (hereafter $TT$ and $TT,TE,EE$, respectively) in combination with the low-$\ell$ \planck\ temperature and polarization APS in the range $2\leq\ell<30$ (lowP). %[\textbf{Likelihood paper 2015}]
%\cite{Likelihood paper 2015}.
%Regarding polarization spectra at high $\ell$, 
We also tested the WMAP APS including both temperature and polarization up to $\ell=1200$.%\cite{Bennett:2013}

%\subsection{Results} \label{sec:results}

\begin{figure*}
\begin{center}
\begin{tabular}{c c}
\includegraphics[width=0.38\columnwidth]{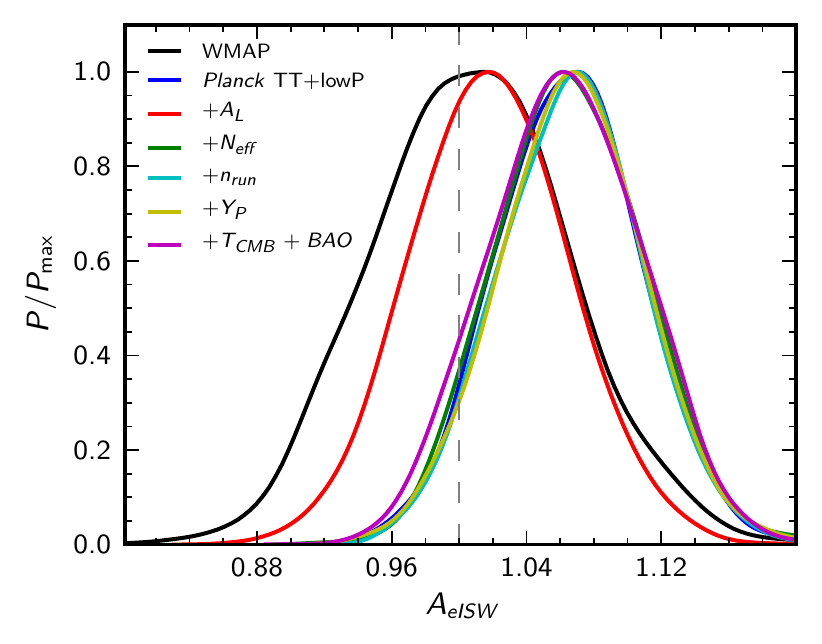}&\includegraphics[width=0.38\columnwidth]{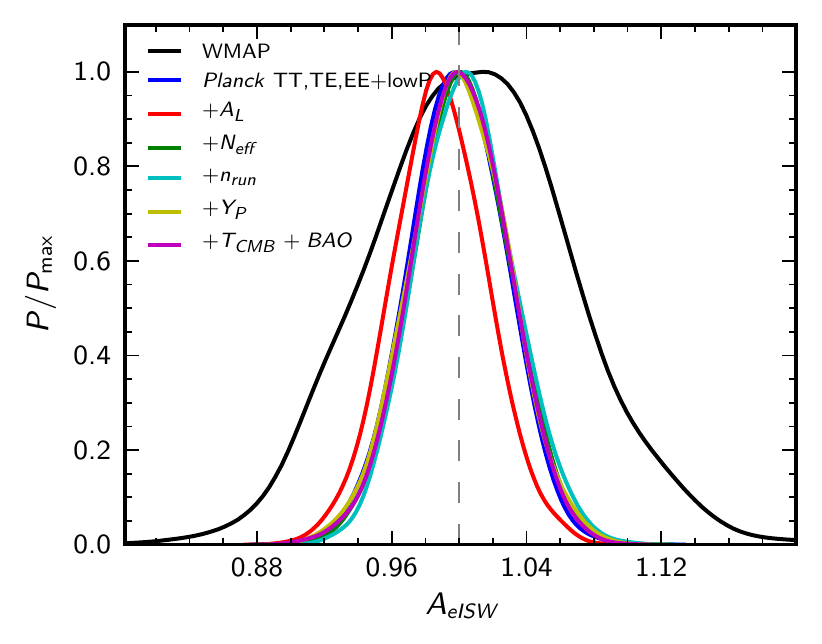}\\
\includegraphics[width=0.38\columnwidth]{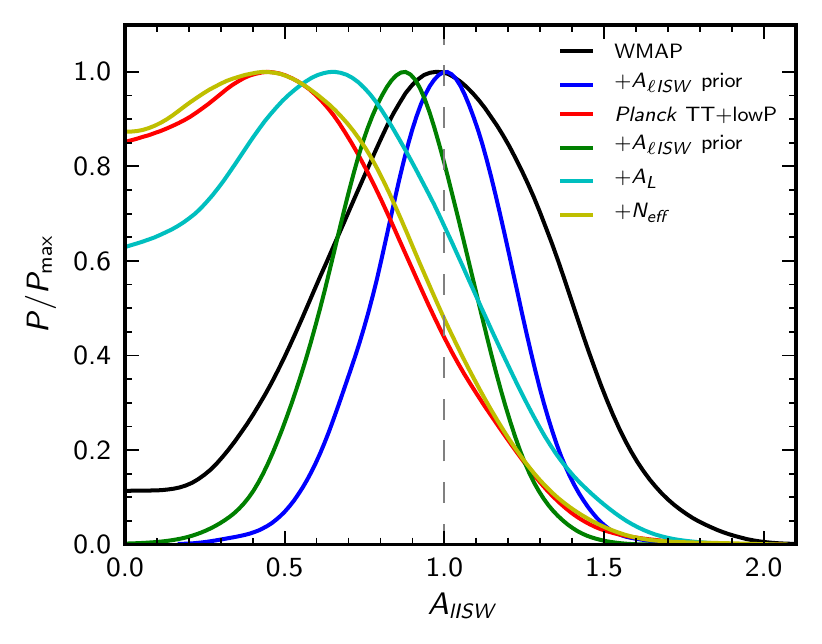}&\includegraphics[width=0.38\columnwidth]{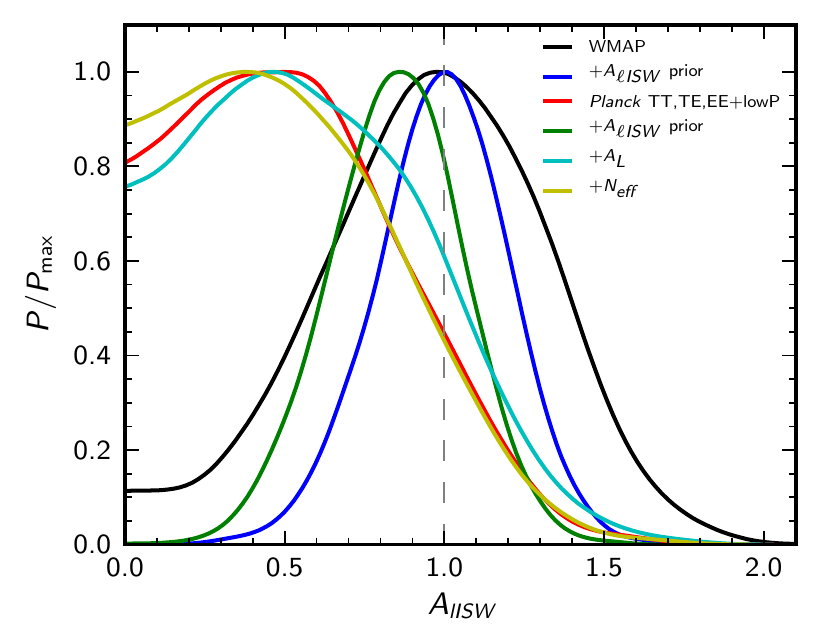}
\end{tabular}
\end{center}
\vskip -0.3cm
\caption{One-dimensional posteriors for $A_{e\text{ISW}}$ %(top panel) 
and $A_{l\text{ISW}}$. % (bottom panel). 
In the left column, only \planck\ temperature and low multipole polarization APS (\planck\ $TT+\text{lowP}$) were used.
The plots in the right column use also the \planck\ polarization APS at high multipoles (\planck\ $TT,TE,EE+\text{lowP}$). From Ref. [25].}
\label{fig:eisw-2D}
\vskip -0.3cm
\end{figure*}

%\noindent %We find that the 
\planck\ $\TT$ data provide tighter constraints than WMAP on the early ISW ($A_{e\text{ISW}} = 1.064^{+0.042}_{-0.043}$ vs. $A_{e\text{ISW}} = 1.007^{+0.056}_{-0.058}$ at 68\% C.L.), and present a $1\sigma$ evidence of $A_{e\text{ISW}}\neq 1$ that is stable when considering the extensions of the $\Lambda$CDM model shown in Fig.~\ref{fig:eisw-2D}.
Regarding the late ISW, \planck\ data place a constraint $A_{l\text{ISW}} < 1.14$ at 95\% C.L.: \planck\ alone does not improve significantly the constraint on $A_{l\text{ISW}}$
with respect to WMAP data (which give $A_{l\text{ISW}} = 0.958^{+0.391}_{-0.317}$ at 68\% C.L.). 
%This is because 
In fact, the late ISW affects 
angular scales that are dominated by cosmic variance, rather than by instrumental noise. Adding the prior on $A_{l\text{ISW}}$ coming from CMB temperature anisotropies-weak lensing correlations, we find a $\sim 4\sigma$ detection $A_{l\text{ISW}} = 0.85\pm0.21$. When we consider the recent \planck\ polarization data at high $\ell$, 
the evidences for a non-standard value of $A_{e\text{ISW}}$ disappear.
Using also the small scale polarization APS does not change the results obtained for $A_{l\text{ISW}}$: their effect is to tighten the upper bounds obtained considering only the $\TT$ APS.

%\begin{figure*}[htb!]

\section{CMB low multipoles anomalies}\label{aba:sec1}

It is usually stated that the six parameters of the $\Lambda$CDM model\cite{Hinshaw:2012aka, Ade:2013zuv,Planck:2015xua} 
are enough to describe the large scale Universe. 
However, some features are not well captured, and anomalies occur for instance at the largest CMB angular scales (see e.g. Ref. [\refcite{Copi:2010na}]), although they are often regarded as mere curiosities. 
We focus here on 
%one of these anomalies, 
the lack of correlation\cite{Hinshaw:1996ut,Spergel:2003cb,Bernui:2006ft,Copi:2006tu,Copi:2008hw,Efstathiou:2009di,Sarkar:2010yj,Gruppuso:2013dba,Copi:2013cya} 
and explain why it deserves attention. The low variance anomaly\cite{Gruppuso:2013xba} 
%\cite{Monteserin:2007fv,Cruz:2010ud,Gruppuso:2013xba,Ade:2013nlj} 
is a closely related observation,\cite{Gruppuso:2013dba} so that 
the terms ``lack of power'' and ``lack of correlation'' are used as synonymous. 

There is a lack of power, with respect to $\Lambda$CDM, in the two-point correlation function of CMB temperature anisotropies for angles $\gsim~60^{\circ}$,
%larger than~$\sim~60^{\circ}$. 
%This intriguing discrepancy was 
as originally noted with COBE\footnote{http://lambda.gsfc.nasa.gov/product/cobe/} data\cite{Hinshaw:1996ut} and then confirmed by the WMAP team already in their first year release.\cite{Spergel:2003cb}
In Ref. [\refcite{Bernui:2006ft}] this feature was associated to missing power in the quadrupole. WMAP-3yrs and WMAP-5yrs data were then used to show\cite{Copi:2006tu, Copi:2008hw} that a lack of correlation occurs only in 0.03\% of the $\Lambda$CDM realizations. A subsequent analysis\cite{Efstathiou:2009di} confirmed the anomaly using WMAP-5yrs data, and, at the same time, found, with a Bayesian approach, that the $\Lambda$CDM model cannot be excluded. WMAP-7yrs data were taken into account 
in Ref. [\refcite{Sarkar:2010yj}], while
WMAP-9yrs data were considered in Ref. [\refcite{Gruppuso:2013dba}], where the lack of correlation was studied against the Galactic masking.
\planck\ 2013 and WMAP-9yrs data were analyzed in Ref. [\refcite{Copi:2013cya}], which confirmed for this anomaly a significance at the level of $99.97\%$.
Similar results were obtained in Ref. [\refcite{Ade:2015hxq}] where \planck\ 2015 data were taken into account.
One intriguing feature of this anomaly is that it is more significant at high Galactic latitude.\cite{Gruppuso:2013dba,Ade:2015hxq} 
Is this a simple statistical fluke or it is caused by a physical mechanism?

We now elaborate on a possible fundamental origin for this effect.\cite{Gruppuso:2015zia,Gruppuso:2015xqa} 
% (see [\refcite{Gruppuso:2015zia,Gruppuso:2015xqa}] and references therein).
Lack of power at large angular scales is a typical manifestation of early departures from slow--roll, which follow naturally the emergence from an initial singularity. As explained in Refs. [\refcite{Dudas:2012vv,Kitazawa:2014dya}], when this occurs the PS approaches in the infrared the limiting behavior
\begin{equation}
{\cal P}(k) = A \, \frac{k^{3}}{(k^2 + \Delta^2)^{2-n_s/2}} \ ,
\label{primordialmodified}
\end{equation}
which brings along a new physical scale $\Delta$. An infrared depression of the 
%APS 
PS presents itself naturally in string theory, %\cite{stringtheory}, 
in orientifold vacua %\cite{orientifolds} 
with high--scale supersymmetry breaking (see e.g. references in Refs. [\refcite{Gruppuso:2015zia,Gruppuso:2015xqa}]).
%\cite{bsb}. 
In these models a scalar field emerges at high speed from an initial singularity, to then bounce against a steep exponential potential before attaining an eventual slow--roll regime. The key ingredient is the steep exponential, whose logarithmic slope is predicted by string theory,\cite{dks} 
%\cite{dks,as13,fss},
and a number of exactly solvable systems provide explicit realizations of this peculiar dynamics.
% \cite{fss}.
%
%\begin{figure}[ht]
\begin{figure}
\centering
\includegraphics[width=60mm]{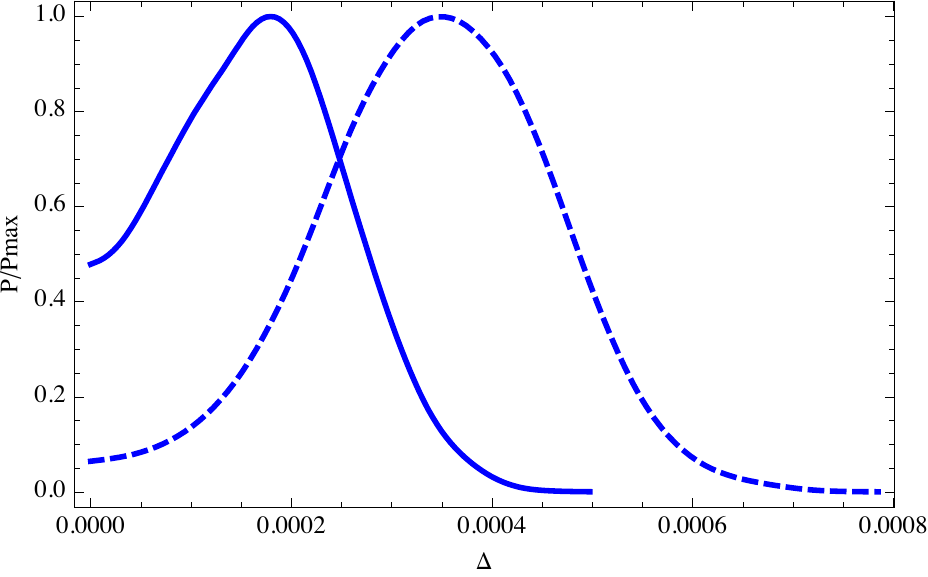}
\caption{\small Posterior probabilities of $\Delta$ (solid line for the standard mask with $f_{sky}~\simeq~90\%$, and dashed line for an extended mask with $f_{sky}~\simeq~40\%$).}
\label{figura2}
\vskip -0.3cm
\end{figure}
The results of a Bayesian analysis extended to all standard cosmological parameters and based on \planck\ 2013 data are shown in Fig.~\ref{figura2},
where posteriors for $\Delta$ are given for two choices of the Galactic mask
%: the standard mask, with $f_{sky} \simeq 90 \%$ (solid line), and an extended mask with $f_{sky} \simeq 40 \%$
(even if equivalent, for an updated and complete analysis see Ref. [\refcite{Gruppuso:2015xqa}]).
%The estimated value of the new scale $\Delta$ for the latter choice is
For the latter choice, the estimated value is
\begin{equation}
\Delta = (0.340 \pm 0.115) \times 10^{-3} \, {\rm Mpc}^{-1}  \, .
\label{Deltalargemask}
\end{equation}
Interestingly, $\Delta$ in eq.~(\ref{Deltalargemask}) is found to differ from $0$ at $99\%$ C.L. and its magnitude appears reasonable.\cite{Gruppuso:2015zia,Gruppuso:2015xqa} 
Moreover, in analogy with the lack of power anomaly, Fig.~\ref{figura2} shows that the significance of this result increases sizably 
for a larger Galactic mask.
%when a larger Galactic mask is used.

In conclusion, the considerations in Refs. [\refcite{Dudas:2012vv,Kitazawa:2014dya}], inspired by string theory, and in particular by the supersymmetry breaking mechanism\cite{bsb}
%of [\refcite{bsb}] 
and the related cosmological dynamics,\cite{dks} 
%of [\refcite{dks}], 
provided the original motivation for the present analysis. 
The resulting scenario would associate $\Delta$ to the onset of the inflationary phase. %, and as we have seen our estimates appear compatible with this interpretation. 
Collecting more information on 
low multipoles of CMB APS
%the infrared tail of the 
%APS 
%PS 
might tell us something more definite about how an inflationary regime was originally attained.

\section{Features in the primordial fluctuations}
\label{features}

\begin{figure}
\begin{center}
\includegraphics[width=118pt]{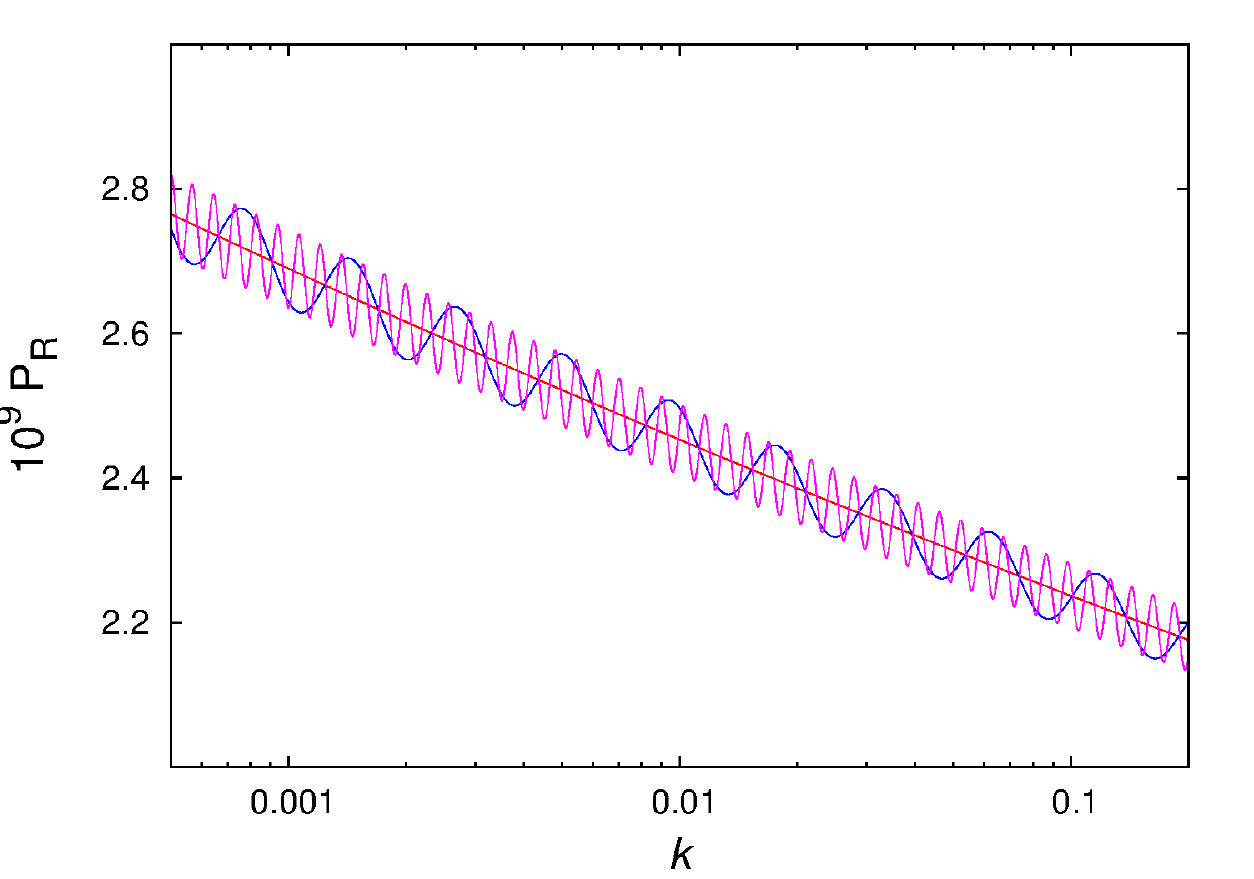}
\includegraphics[width=118pt]{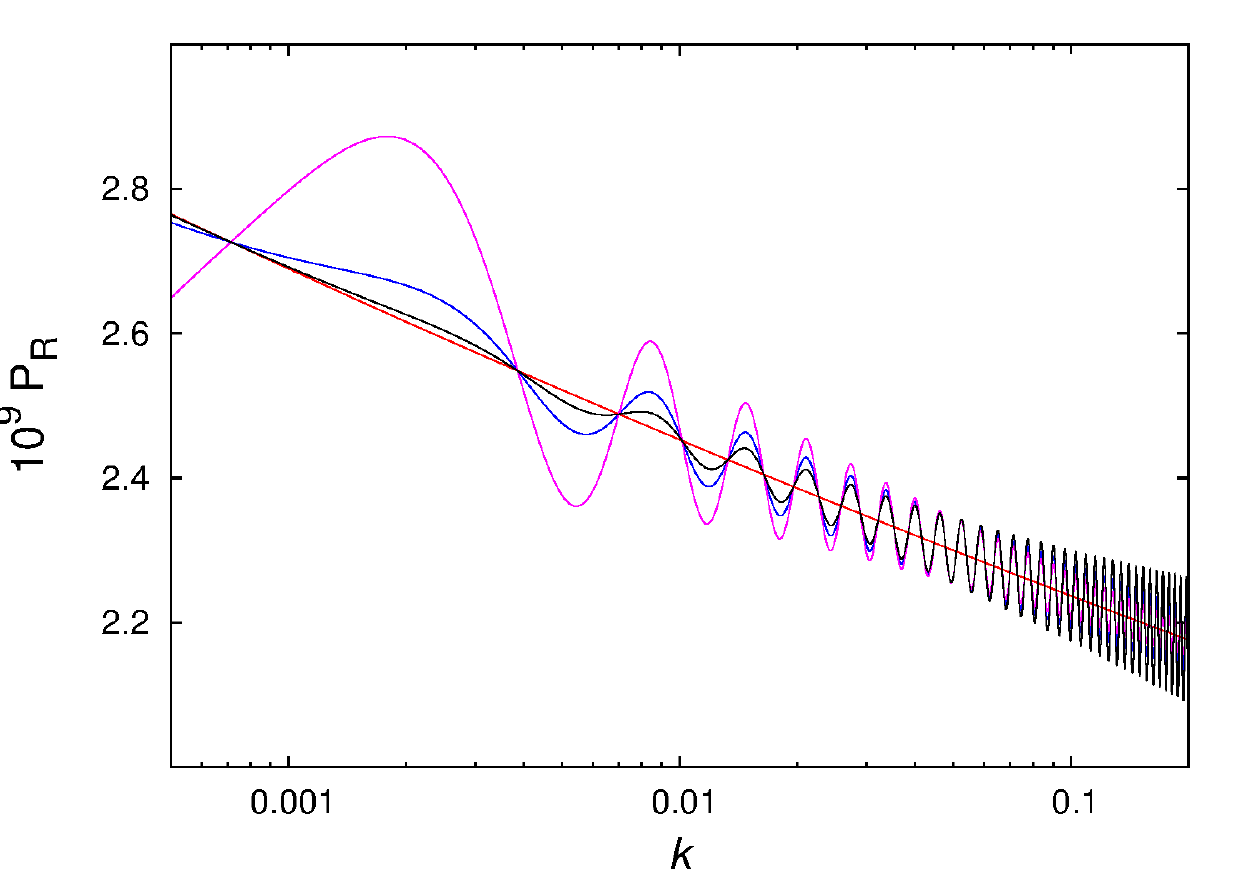}
\includegraphics[width=118pt]{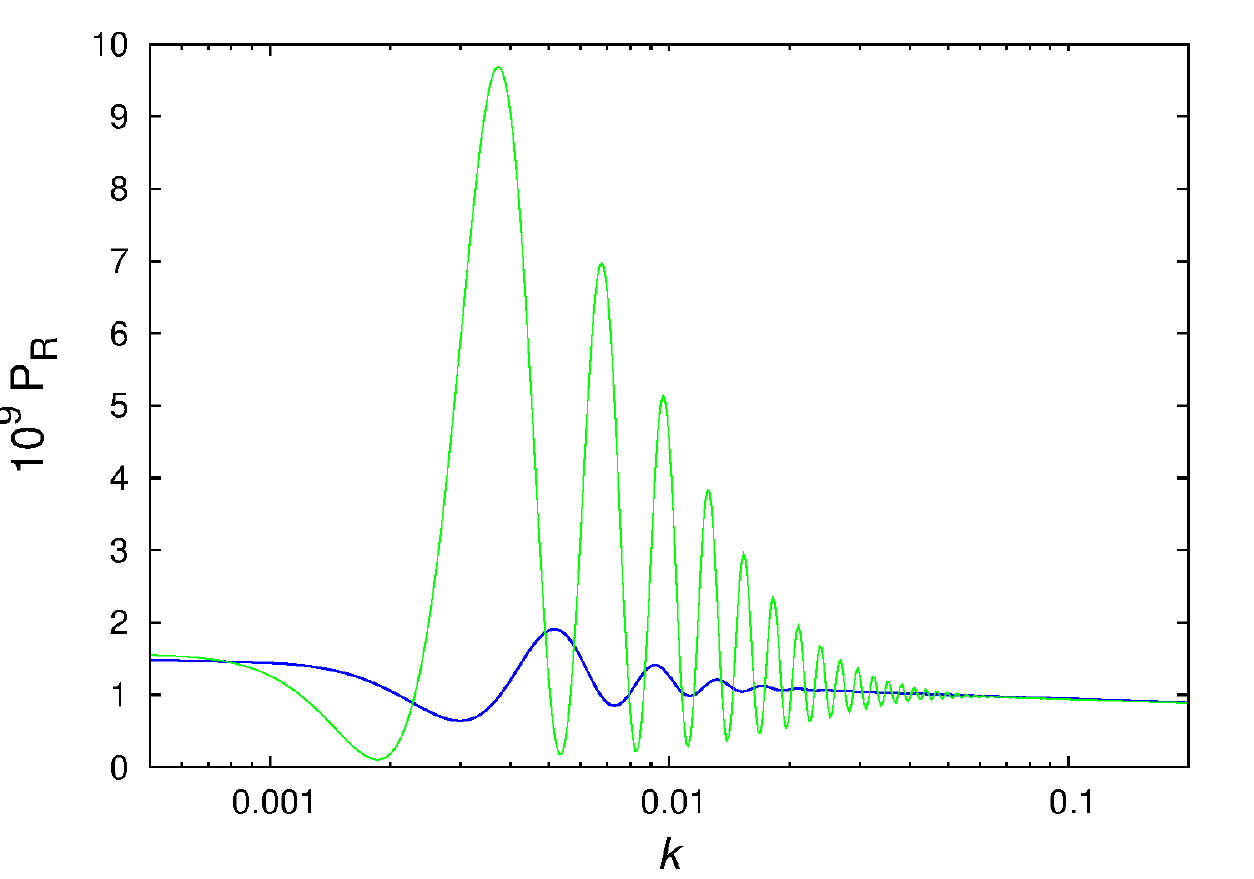}
\end{center}
\caption{PPS simulated for various oscillatory models parametrized with an amplitude, a frequency and a phase 
(for the Linear oscillation models we have the scale-dependence index as further parameter).
Left: Log-spaced oscillations models for different values of the frequency oscillation parameter.
Middle: Linear oscillation models for different values of the scale-dependence index.
Right: Step oscillation models for different values of the frequency and phase parameters. 
}
\label{fig:PR_spectrum}
\end{figure}

The most recent CMB data by the {\it Planck} satellite\cite{Planck:2015xua}
are in excellent agreement with the assumption of adiabatic primordial scalar perturbation with 
nearly scale-invariant PS, described by a simple power law with spectral index $n_s$ 
very close to (albeit different from) unity.\cite{Ade:2015lrj}
It would be produced in the simplest inflationary scenario, in which a single minimally-coupled scalar field slowly rolls down a smooth potential. 
In spite of this, models that account for localized `features' in the PPS
could provide a better fit to data with respect to a smooth power-law spectrum. 
These features could be produced in inflationary models with departures from the near-scale-invariant-power-law spectrum of the standard simplest scenario, and %the 
observable signatures would be in the CMB anisotropy temperature APS and in the matter PS from galaxy surveys. We analyze here three classes of models, and
Fig. \ref{fig:PR_spectrum} displays the corresponding PPS shapes.\\
%%%%%%%%%%%%%%%%%%%%%%%%%%%%%%%%%
%%%%%%FIGURE%%%%%%%%%%%%%%%%%%%%%
%%%%%%%%%%%%%%%%%%%%%%%%%%%%%%%%%%%%%%
%%%%%%%%%%%%%%%%%%%%%%%%%%%%%%%%%%%%%%%
%The inflationary class of models we analyze are:
%, namely:
%
(i) Log-spaced oscillations model assumes an oscillation in proper time affecting the amplitude of curvature perturbation during the inflationary expansion (producing features periodic in $\ln k$).
It is the case of models with no-Bunch-Davies initial condition,\cite{Danielsson:2002kx} or in the bouncing inflationary scenario\cite{Biswas:2010si} or also in the axion monodromy inflation.\cite{Flauger:2009ab} 
%The shape of the PPS is showed in the top-left panel of Fig. \ref{fig:PR_spectrum}.
%
(ii) Linear oscillation model includes effects from a possible boundary on effective field theory (where we assume new physics which occurs at one moment in time, such as a discontinuity
in single-field inflation\cite{Starobinsky:1992ts} or a sharp turn in multi-field inflation\cite{Shiu:2011qw}).
%In the top-right panel of Fig. \ref{fig:PR_spectrum} we show the PPS simulations for this kind of models.
%
(iii) Step oscillation model assumes a briefly interruption of the slow-roll. 
For instance it can happen by a phase transition, a burst of resonant particle production, a sudden turn in field space or a step in the inflation potential.\cite{Adams:2001vc} 
%Looking at the bottom panel of the Fig. \ref{fig:PR_spectrum}, we can see the spectrum of primordial perturbations for these models. 
It is found to be essentially a power-law with superimposed oscillations localized only in a limited range of wavenumbers.
It is noteworthy that this kind of models is able to produce oscillation at very high-$\ell$, and it is very interesting looking to the CMB temperature anisotropies glitches in correspondence of $\ell = 22$ and $\ell=40$, first observed by the WMAP experiment and later confirmed by the {\it Planck} satellite.

We used here
the {\it Planck} 2013 data release and a combined CMB and SDSS (DR-11)\cite{Beutler:2013yhm} dataset. 
Our results show that using the CMB data alone we have no-evidence of improving the concordance with data and agree with the more recent results of the {\it Planck} Collaboration.\cite{Ade:2015lrj}
Instead, using the combined dataset of CMB SDSS-DR11 data, we can see a positive Bayesian evidence for the inflationary log-spaced oscillation and step oscillation models.
Updating this analysis with final {\it Planck} 
%and SDSS-DR11 
data will be very interesting 
%in order 
to confirm or discard these kinds of models.
%The updated results of our analysis will use the {\it Planck} (2015) and SDSS-DR11 data and will be very interesting in order to confirm or discard these kinds of models.

\def\Journal#1#2#3#4{{#1} {\bf #2}, #3 (#4)}

%%%%%%%%%%%%%%%%%%%%%%%%%%%%%%%%%%%%%%%%%%%%%%%%%%

\section{Primordial non-Gaussianity}
\label{nongauss}

The study of primordial non-Gaussianity (PNG) provides a powerful tool to shed light on early Universe mechanisms for the generation of primordial 
perturbations (see e.g. Refs. [\refcite{2004PhR...402..103B,2010AdAst2010E..72C}] and references therein).  Different inflationary models predict different {\it amplitudes}, {\it shapes}, and {\it scale dependence} of PNG. As a result, PNG allows to discriminate between %inflationary 
models that can show degeneracies 
considering only the APS. 
%on the basis of an APS-only analysis. 

One of the main goals of these analyses 
is to constrain the amplitude and shape of PNG using the angular bispectrum of CMB anisotropies,
%The CMB angular bispectrum is 
related via {\em linear} radiation transfer to the primordial bispectrum, $B_{\Phi}(k_1,k_2,k_3)$, defined by
\begin{equation}
\label{bispectrumPhi}
\langle  \Phi({\vec k}_1) \Phi({\vec k}_2) \Phi({\vec k}_3) \rangle= (2 \pi)^3 \delta^{(3)}({\vec k}_1+{\vec k}_2+{\vec k}_3) B_{\rm \Phi}(k_1,k_2,k_3) \,;
\end{equation}
here $\Phi$ is the primordial potential defined in terms of the comoving curvature perturbation $\zeta$ on super-horizon scales by $\Phi \equiv (3/5) \zeta$. 
The bispectrum %$B_{\Phi}(k_1,k_2,k_3)$ 
measures the correlation among three perturbation modes, and it is expected to be zero for Gaussian perturbations. In general, 
the bispectrum can be written as
\begin{equation}
\label{amplitudeandF}
B_{\Phi}(k_1,k_2,k_3)= f_{\rm NL} F(k_1,k_2,k_3) \, ,
\end{equation}
where $f_{\rm NL}$ is the so-called ``nonlinearity parameter'', a dimensionless parameter measuring the amplitude of non-Gaussianity.
The functional dependence of $F(k_1,k_2,k_3)$ on the type of triangle formed by $k_1, k_2, k_3$ defines the \emph{shape} of the bispectrum.\cite{2004JCAP...08..009B}
Even in the simplest models of inflation, consisting of a single slowly-rolling scalar field, some level of PNG is predicted,\cite{2003NuPhB.667..119A,2003JHEP...05..013M}
but this is too small to be detectable in CMB and LSS surveys. Large level of PNG can be produced however in multi-field scenarios, or in single-field models with non-standard 
Lagrangians and deviations from Bunch-Davies vacuum, and in many other cases. Each of the scenarios outlined above predicts different shapes, the main of which are briefly described below.

\noindent {\em Local shape}, where the signal peaks in ``squeezed''  triangles ($k_1 \ll k_2\simeq k_3$). This shape is typically generated in multi-field models of inflation.
\noindent {\em Equilateral shape}, peaking on equilateral bispectrum triangles($k_1 \approx k_2 \approx k_3$). Examples of this class include single-field models with non-canonical
kinetic term,\cite{2007JCAP...01..002C} such as e.g. Dirac-Born-Infeld (DBI) inflation,\cite{2004PhRvD..70j3505S} models characterized by more general higher-derivative interactions of the inflaton field, and models arising from effective field 
theories.\cite{2008JHEP...03..014C}
\noindent {\em Folded (flattened) shape}, peaking on isosceles, nearly degenerate triangles.
Examples of this class include e.g. single-field models with non-Bunch-Davies vacuum. 
\noindent{\em Orthogonal shape},
which is generated, e.g., in single-field models of inflation with a non-canonical kinetic term, or with general higher-derivative interactions. Notably,  the folded shape described above can 
be obtained as a linear combination of equilateral and orthogonal shapes. In light of this, actual measurements of PNG generally focus on  local, equilateral and orthogonal templates.

It must be noted that many but not all models are included in the previous classification. For example, models with a temporary breaking of slow-roll conditions generate strongly scale-dependent, oscillatory shapes that cannot be approximated by a combination of local, equilateral and orthogonal templates. Given the limited scope of this review, the focus here will be however only on the above main shapes.

Having computed the CMB angular bispectrum templates arising from the various primordial shapes, non-Gaussianity estimation consists then essentially in fitting such templates to the 3-point function of the data in order to measure the best-fit amplitude parameters $f_{\rm NL}$. This apparently straightforward approach actually presents many statistical and numerical complications. These arise mainly from the huge number of modes 
(triangles) which contribute to the signal for high resolution experiments like {\it Planck} and WMAP, and from spurious mode couplings from 
sky-cut and anisotropic instrumental noise.  Several numerical techniques have been successfully tested in the literature and implemented by the {\it Planck} team in order to address these issues. This gave rise to several independent, but ultimately equivalent, {\it Planck} bispectrum analysis pipelines, the so called modal, binned and separable template-fitting
%(KSW) 
estimators, that were separately applied to the data.\cite{2005ApJ...634...14K,2010MNRAS.401.2406M,2010PhRvD..82b3502F,2012JCAP...12..032F,2010MNRAS.407.2193B} We refer the reader 
to {\it Planck} papers on PNG,\cite{Planckpaper,Planckpaper2015} for all details of the analysis, including 
%the large battery of 
validation tests 
%that were 
carried on data and simulations, 
%an 
%accurate 
descriptions of the various pipelines, and constraints on a much larger set of shapes than the three discussed here. 

The final \planck\ 
 results for the local, equilateral and orthogonal shapes, from the 2015 combined analysis of temperature and polarization data, are as follows:
%\begin{eqnarray}\label{tab:fnlresults}
%f_{\rm NL}^{\rm local} \, & =& +0.8\pm 5.0 \nonumber \\
%f_{\rm NL}^{\rm equil.}  &=&  -4 \pm 43 \nonumber \\
%f_{\rm NL}^{\rm ortho.} &=&  -26 \pm 21 \; .
\begin{equation}
%\label{tab:fnlresults}
%f_{\rm NL}^{\rm local} \, & =& +0.8\pm 5.0 \; ; \;\; f_{\rm NL}^{\rm equil.}  &=&  -4 \pm 43 \; ; \;\; f_{\rm NL}^{\rm ortho.} &=&  -26 \pm 21 \; .
f_{\rm NL}^{\rm local} \,  = +0.8\pm 5.0 \; ; \;\; f_{\rm NL}^{\rm equil.}  =  -4 \pm 43 \; ; \;\; f_{\rm NL}^{\rm ortho.} =  -26 \pm 21 \; .
\end{equation}

The main conclusion from {\it Planck} %PNG analysis 
is that consistency with Gaussianity is found in all cases (including shapes not considered here). {\it Planck} bispectrum constraints lead to important 
implications for inflationary model building, such as a lower bound on the sound speed in effective single field inflation theory, or 
limits on the curvaton decay fraction, and so on. In light of the current results, the simplest slow-roll single field inflationary paradigm has passed its most stringent and accurate test to date (although alternative, more complex, possibilities, while constrained, are by no means ruled out yet).
{\it Planck} has extracted nearly all the possible PNG information from CMB data. Even an ideal, noiseless temperature$+$polarization experiment would improve on current 
error bars by at most a factor $2$. For substantial improvements it will be necessary to look at different observables and wait for future experiments. LSS could be in principle 
promising, since it contains more modes than the CMB. Precise primordial bispectrum estimation from LSS surveys is however very hard due to non-linearities from gravitational evolution, galaxy-bias and other effects; whether we will be able to achieve improvements from the LSS bispectrum will strongly depend on how well we can keep all these systematics under control, and it is at present an open question. Two-point function based measurements of large scale signatures arising from scale-dependent halo bias look on the other hand quite promising, and have the potential 
of achieving $f_{\rm NL} \sim 1$ sensitivity for the local shape.\cite{2012MNRAS.422.2854G} 
It has been pointed out that the study of the bispectrum of 21 cm radiation or measurements of cross-correlations between CMB spectral distortions and temperature anisotropies can in principle improve over current bounds by more than one order of magnitude. These are fascinating but futuristic prospects, since high-sensitivity $f_{\rm NL}$ measurements with these techniques will  require either full-sky 21 cm surveys with redshift tomography in the  $30 \lesssim z \lesssim 100$ range, or high-resolution maps of angular anisotropies of the CMB $\mu$-distortion parameter.\cite{Munoz:2015eqa,2012PhRvL.109b1302P}

\renewcommand{\arraystretch}{1.5}

%neutrino masses and mixing angles
\newcommand{\Dm}{\Delta m^2_{31}}
\newcommand{\dm}{\Delta m^2_{21}}
\newcommand{\Dmsol}{\Delta m^2_\mathrm{sol}}
\newcommand{\Dmatm}{\Delta m^2_\mathrm{atm}}
\newcommand{\ths}{\theta_{12}}
\newcommand{\tha}{\theta_{23}}
\newcommand{\thb}{\theta_{13}}
\newcommand{\mb}{m_\beta}
\newcommand{\mbb}{m_{\beta\beta}}

\newcommand{\calP}{\mathcal{P}}
\newcommand{\calL}{\mathcal{L}}
\newcommand{\bsy}{\boldsymbol}
\newcommand{\bth}{\bsy{\theta}}
\newcommand{\bd}{\bsy{d}}
\newcommand{\thalf}{T_{1/2}^{0\nu}}
\newcommand{\Nobs}{N_\mathrm{obs}}
\newcommand{\Tri}{^3\mathrm{H}}
\newcommand{\Ge}{^{76}\mathrm{Ge}}
\newcommand{\Xe}{^{136}\mathrm{Xe}}
\newcommand{\nsig}{N_\mathrm{s}}
\newcommand{\nbkg}{N_\mathrm{b}}
\newcommand{\nme}{\mathcal{M}^{(0\nu)}}
\newcommand{\nmef}{\mathcal{M}_\mathrm{fid}^{(0\nu)}}
\newcommand{\nldb}{0\nu2\beta}

\newcommand{\ueV}{\mu\mathrm{eV}}
\newcommand{\yr}{\mathrm{yr}}

% placeholder 
\newcommand{\plhold}[1]{\textbf{[#1]}}

% it renames \sqrt as \oldsqrt
\let\oldsqrt\sqrt
% it defines the new \sqrt in terms of the old one
\def\sqrt{\mathpalette\DHLhksqrt}
\def\DHLhksqrt#1#2{%
\setbox0=\hbox{$#1\oldsqrt{#2\,}$}\dimen0=\ht0
\advance\dimen0-0.2\ht0
\setbox2=\hbox{\vrule height\ht0 depth -\dimen0}%
{\box0\lower0.4pt\box2}}

\section{Limits on neutrino masses from cosmology and particle physics}
\label{neutrinossec}

The absolute scale of neutrino masses is one of the main open issues both in cosmology and particle physics. Current experimental strategies involve i) measurements exploiting kinematics effects in beta
decay:\cite{Drexlin:2013lha}, ii) searches for neutrinoless double beta decay ('$\nldb$'),\cite{Cremonesi:2013vla} and iii) cosmological observations.\cite{Lesgourgues:2006nd} The three approaches are complementary, each of them presenting its own advantages and disadvantages and being sensitive to slightly different quantities related to the neutrino masses.\cite{joint} In this work, we derive joint constraints on neutrino mass parameters from the most recent observations from both laboratory and cosmological experiments, and forecasts, combining them in the framework of Bayesian statistics. In particular, for '$\nldb$' experiments, we take into account the uncertainty related to nuclear matrix elements, in order to account its impact on the neutrino mass estimates.

\subsection{Neutrino parameters, method and data}

We denote the masses of the neutrino mass eigenstates $\nu_i$ with $m_i$ $(i=1,\,2,\,3$). $\dm$ represents the difference between the two eigenstates closest in mass, while the sign of $\Dm$ discriminates between the normal (NH, $\Dm > 0$) and inverted (IH, $\Dm < 0$) hierarchies.
The neutrino mass eigenstates are related to the flavour eigenstates $\nu_\alpha$ ($\alpha=e,\,\mu,\,\tau$) through
$\nu_\alpha = \sum_i U_{\alpha i} \nu_i$, 
where $U_{\alpha i}$ are the elements of the neutrino mixing matrix $U$,
parameterized by the three mixing angles $(\ths,\,\tha,\,\thb)$, one Dirac ($\delta$) and two Majorana ($\alpha_{21},\,\alpha_{31}$) CP-violating phases.
Oscillation phenomena are
insensitive to the %two 
Majorana phases, that however affect $0\nu2\beta$ decay.
The different
combinations of the mass eigenvalues and of the elements of the mixing matrix probed by the experimental avenues are: %the following: 
the squared effective electron neutrino mass
$\mb^2 \equiv  \sum_i \left| U_{ei}\right|^2m_i^2$ ($\beta$ decay
experiments), the effective Majorana mass
$\mbb\equiv \left| \sum_i U_{ei}^2 m_i\right|$ ($0\nu2\beta$ searches),
the sum of neutrino masses
$M_\nu \equiv \sum_{i=1}^{3} m_i$ 
%= m_1+m_2+m_3$ 
(cosmological observations).
%%%%%%%%%%%%%%%%%%% METHOD %%%%%%%%%%%%%%%%%%
We perform a Bayesian analysis based on a MCMC method,
using \texttt{cosmoMC}\cite{Lewis:2002ah} as a generic sampler.
We consider the following vector of base parameters:
$\left(M_\nu,\,\dm, \Dm,\,\sin^2 \ths,\,\sin^2 \thb,\,\phi_2,\,\phi_3,\,\xi \right)$
where $\phi_2 \equiv \alpha_{21}$, $\phi_3 \equiv \alpha_{31} - 2\delta$ and
$\xi$ is a ``nuisance'' parameter related to the nuclear modeling uncertainty. 
%(see below). 
We assume uniform prior distributions for all parameters
%. We do not consider 
and neglect the mixing angle $\theta_{23}$, irrelevant for mass parameters. 
%since 
%%none of the 
%mass parameters 
%do not 
%depend on it. 

Our baseline dataset is the global fit of the updated neutrino oscillation parameters.\cite{Forero:2014bxa} 
%updated after the Neutrino 2014 conference. 
We model the likelihood as a the product of individual Gaussians (correlations 
can be neglected\cite{Forero:2014bxa,GonzalezGarcia:2012sz}).\footnote{See {http://www.nu-fit.org/} for results updated after the Neutrino 2014 conference.} 
%\footnote{See \url{http://www.nu-fit.org/} for results updated after the Neutrino 2014 conference.}
%For direct measurements, we consider 
KATRIN\cite{Osipowicz:2001sq} and HOLMES\cite{Alpert:2014lfa} 
represent 
our forthcoming and next-generation direct measurement datasets, 
respectively.
We take the likelihood for kinematic measurements to be a Gaussian in $\mb^2 > 0$,
with a width given by the expected sensitivity of the experiment, i.e. $\sigma(\mb^2) = 0.025,\, 0.006\, {\rm eV}^2$ for KATRIN and HOLMES, respectively.
For $0\nu2\beta$ searches, we consider the current data from the GERDA experiment\cite{Agostini:2013mzu} as the present dataset,
its upgrade (GERDA-II) for the near-future, and the nEXO experiment\footnote{https://www-project.slac.stanford.edu/exo/}
as a next-generation dataset. 
$0\nu2\beta$ experiments are sensitive to the half-life of $0\nu2\beta$ decay $\thalf$. If neutrinos are Majorana particles, $\thalf$ is
related to the Majorana effective mass through
%%%
$\thalf = \frac{1}{G^{0\nu}\left|\mathcal{M}^{0\nu}\right|^2}\frac{m_e^2}{\mbb^2}$,
%%%%
where $m_e$ is the electron mass, $G^{0\nu}$ is a phase space factor and $M^{0\nu}$ is the nuclear matrix element. 

We model the likelihood of $0\nu2\beta$ experiments as a Poisson distribution in the number of observed events, with an 
expected value $\lambda = \lambda_S + \lambda_B$ given by the sum of signal ($S$) and background  ($B$) contributions. For a given value of $\thalf$, the expected number of signal events observed in a time $T_\mathrm{obs}$ for a detector mass $M$ is
$\lambda_S=\frac{\ln 2 N_A \mathcal{E} \epsilon}{m_{enr} \thalf}$
where $N_A$ is Avogadro's number, $\mathcal{E}\equiv M T_\mathrm{obs}$ is the exposure, $\epsilon$ is the detector efficiency, $m_{enr}$ is the molar mass of the enriched element involved in the decay. The level of background is given by the ``background index'', i.e. the number
of expected background events per unit mass and time within an energy bin of unit width. For GERDA-I, we use the parameters 
reported in Tab. I of Ref. [\refcite{Agostini:2013mzu}] for the case with pulse-shape discrimination. For GERDA-II, we consider
a reduction of the background index down to $10^{-3}\,\mathrm{counts}\, {\rm keV}^{-1}\mathrm{kg}^{-1}\yr^{-1}$, 
a total exposure of 120 kg yr, and the same efficiency as GERDA-I.\cite{priv_catta} For nEXO, we assume a background index corresponding 
to 3.7 events $\mathrm{ton}^{-1}\yr^{-1}$ in the region of interest and an exposure of 25 ton yr,\cite{NOWpocar} and the same
efficiency as EXO.\cite{Albert:2014awa}

In order to account for the uncertainty related to nuclear modeling,\cite{nuclear} 
we compute $\thalf$ for a given $\mbb$ using fiducial values of nuclear matrix elements (NME) and axial coupling constant, and then rescale it
by a factor $\xi^2$ (see Ref. [\refcite{Minakata:2014jba}] for a similar approach). 
For what concerns the cosmological dataset,
we use the posterior distribution of $M_\nu$ from the combination of \planck\ temperature and polarization data with baryon acoustic oscillations (BAO),\cite{Planck:2015xua} 
as both our current and forthcoming reference dataset. 
Finally, we consider the {\it Euclid} mission (weak lensing tomography, galaxy clustering and ISW) in combination with data from {\it Planck}\cite{Laureijs2011} 
%\cite{Laureijs:2011gra} 
as our reference next-generation experiment. 
We model the likelihood as Gaussian in $M_\nu=0.1\, {\rm eV}$, with
$\sigma(M_\nu)=0.06 \, {\rm eV}$ and the addition of the physical prior $M_\nu >0$. 

%\begin{figure*}[htb!]
\begin{figure*}
\begin{center}
\begin{tabular}{c c c}
\includegraphics[width=0.3\textwidth]{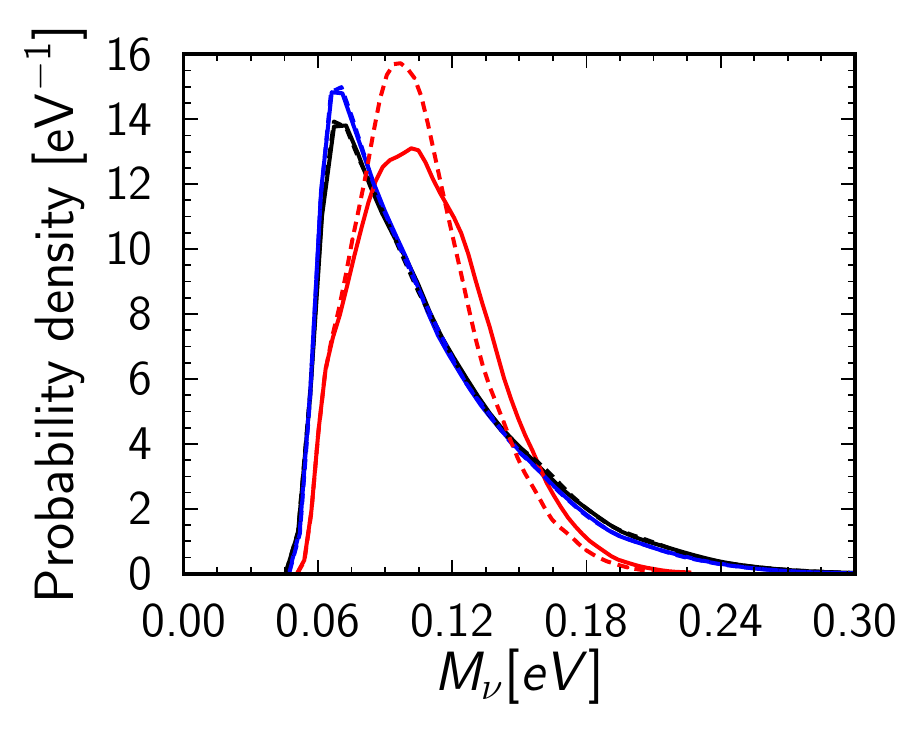} & 
\includegraphics[width=0.3\textwidth]{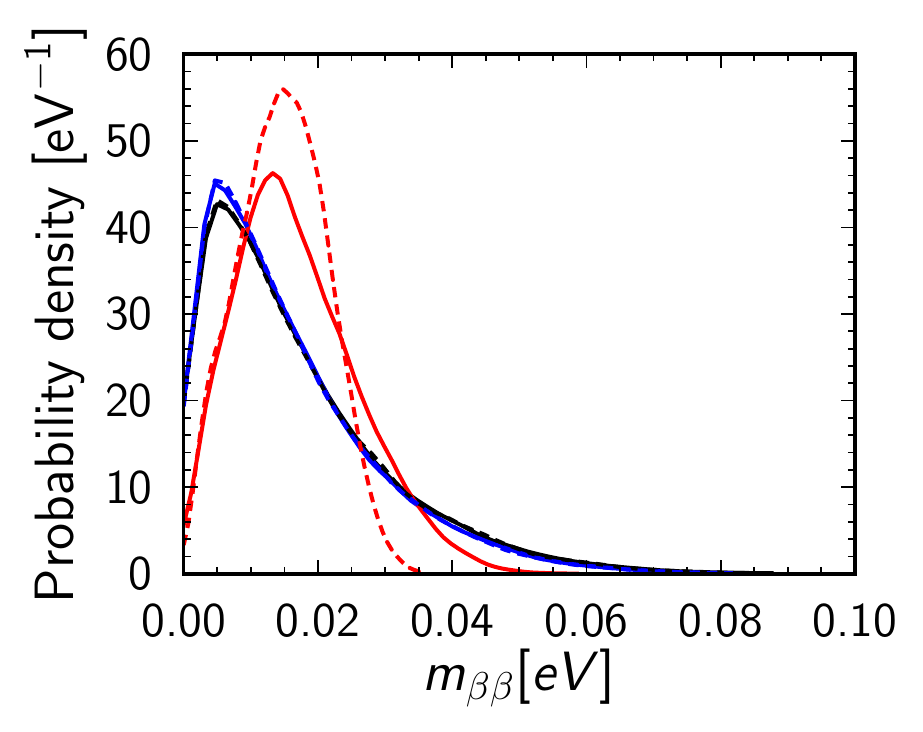} &
\includegraphics[width=0.3\textwidth]{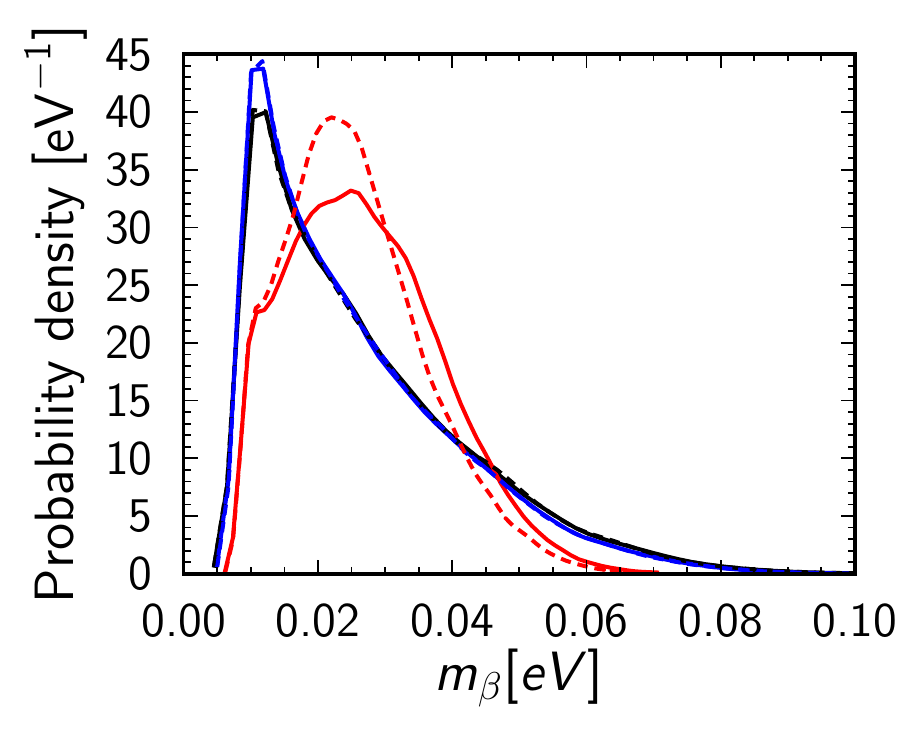} \\
\includegraphics[width=0.3\textwidth]{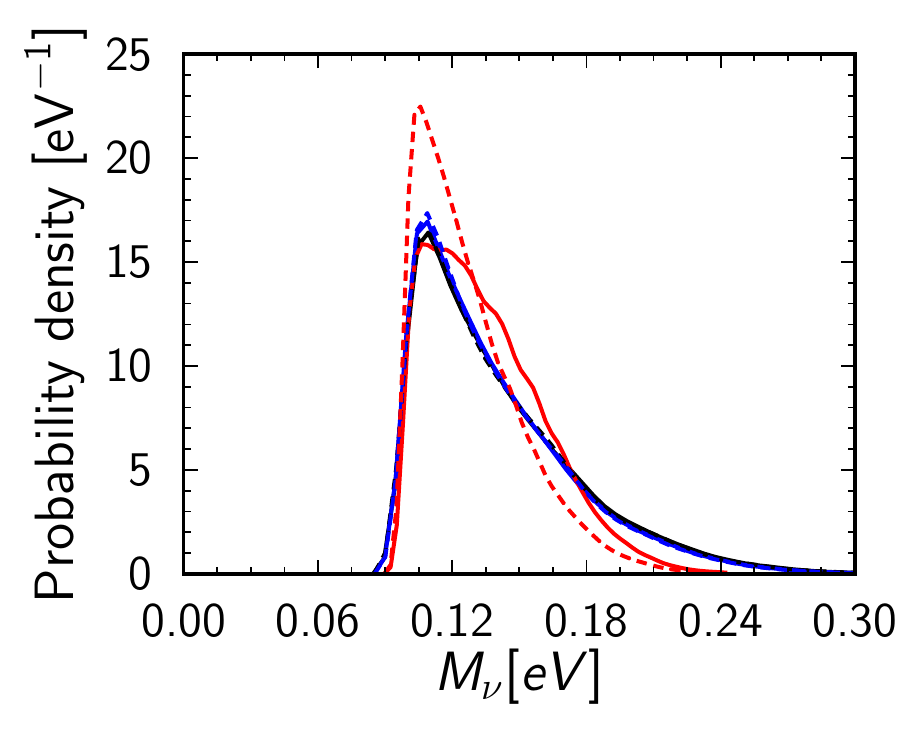} & 
\includegraphics[width=0.3\textwidth]{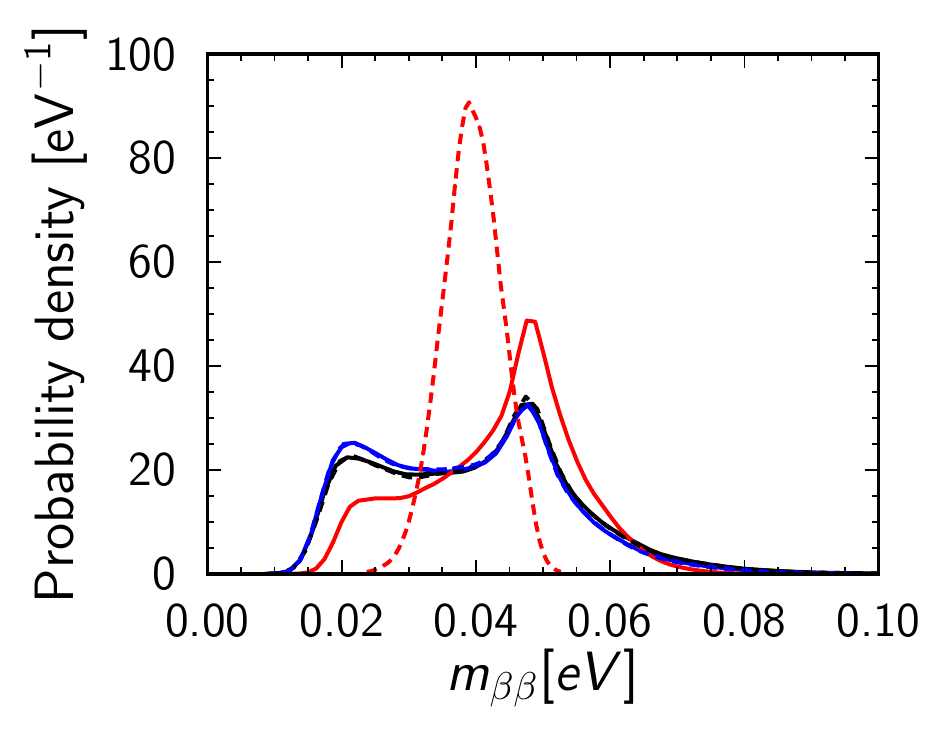} &
\includegraphics[width=0.3\textwidth]{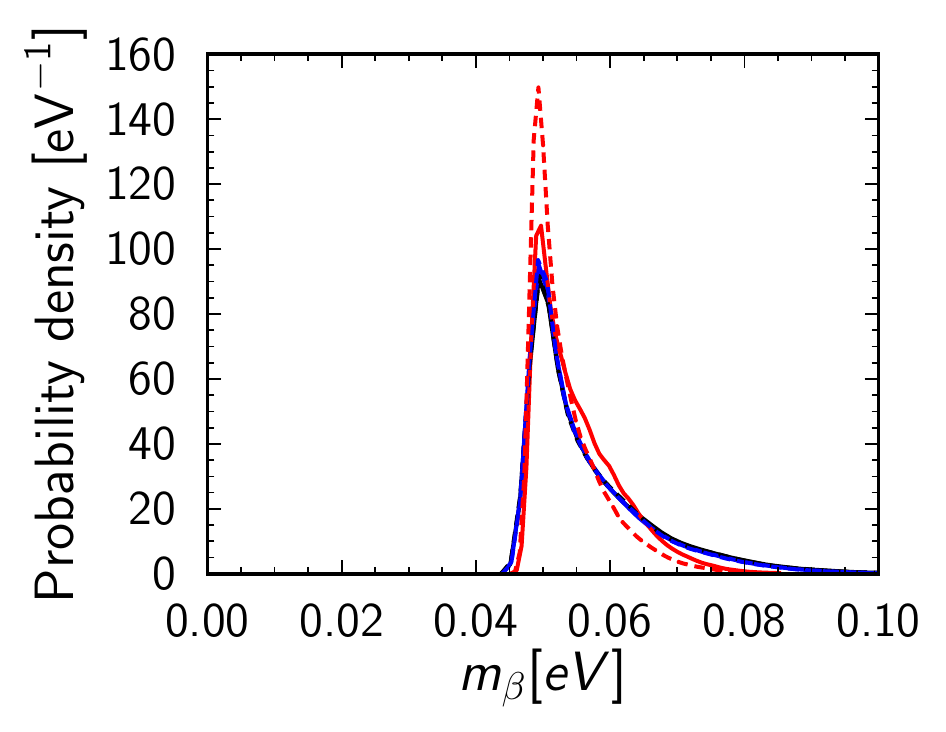} 
\end{tabular}
\end{center}
\vskip -0.3cm
\caption{Posterior distributions for the neutrino mass parameters, for NH (top row) and IH (bottom row). 
Solid (dashed) curves correspond to marginalization over nuclear uncertainties (fixed fiducial values for nuclear parameters). Black, blue and red curves refer to present, forthcoming and next-generation datasets, respectively. }
\label{fig:post1D}
\vskip -0.3cm
\end{figure*}

\subsection{Results} 

We present our results for $M_\nu$, $\mb$ and $\mbb$ in Fig. \ref{fig:post1D}, both in the case where $\xi$ is fixed to 1 and when $\xi$ is marginalized over, in order to show the
impact of uncertainties in nuclear modeling. 
Notice that the low mass region
is excluded by the oscillation data, with the only exception of $\mbb$ in the case of NH; the reason
is that in this case the phases can arrange in order to yield $\mbb=0$ even for finite values of the
mass differences. Similar limits are provided by the ``present'' dataset
independently of whether nuclear uncertainties are marginalized over: present constraints 
are dominated by the cosmological limit on $M_\nu$, that translates directly to bounds on $\mb$ and $\mbb$ once
oscillation data are taken into account. 
%A similar conclusion can be drawn by noticing that the direct limits on these parameters are much weaker. 
Forthcoming datasets yield similar constraints for the mass parameters: the upgraded sensitivity of GERDA-II and
the inclusion of KATRIN provide a marginal improvement to the \planck +BAO plus oscillations data combination.
Substantial differences arise for next-generation experiments. In this case, cosmological observations
and $0\nu2\beta$ searches have comparable constraining power, and
the nuclear uncertainties have a dramatic impact in deriving parameter constraints.
Marginal evidence for non-minimal mass parameters can be highlighted in the case of normal hierarchy,
even when nuclear uncertainties are taken into account. 

%In conclusion, the 
The combination of current and forthcoming data from oscillation, kinematic, $0\nu2\beta$ and cosmological experiments
allows to put upper bounds on the neutrino mass parameters.
Since these limits are dominated by the combination of oscillations and cosmological data, they are not affected by uncertainties in nuclear modeling.
%Assuming 
For
$M_\nu=0.1\, {\rm eV}$ and a 
factor 2 uncertainty in nuclear modeling, 
%next-generation 
future experiments will ideally allow to
measure non-minimal mass parameters with a 95\% accuracy. 

\newcommand{\mpl}{m_\mathrm{Pl}}
\newcommand{\srp}[1]{\lambda^{(#1)}_\mathrm{H}}
\newcommand{\srpl}{\srp{\ell}}
\newcommand{\fg}{\mathfrak{g}}
\newcommand{\fso}{\mathfrak{so}} 
\newcommand{\SO}{\mathrm{SO}}
\renewcommand{\O}{\mathrm{O}} 
\newcommand{\Cl}{\mathrm{C}\ell}
\newcommand{\Cc}{{\cal C}} 
\newcommand{\Spin}{\mathrm{Spin}}
\newcommand{\SL}{\mathrm{SL}} 
\newcommand{\SU}{\mathrm{SU}}
\newcommand{\RR}{\mathbb{R}} 
\newcommand{\CC}{{\bf C}}
\newcommand{\CP}{\mathbb{CP}}
\newcommand{\KK}{\mathbb{K}}
\newcommand{\VV}{\mathbb{V}} 
\newcommand{\ZZ}{\mathbb{Z}}
\newcommand{\eL}{\mathcal{L}} 
\newcommand{\eO}{\mathcal{O}}
\newcommand{\AdS}{AdS} 
\newcommand{\dvol}{dvol}
\newcommand{\be}{\boldsymbol{e}} 
\newcommand{\bv}{\boldsymbol{v}}
\newcommand{\bx}{\boldsymbol{x}} 
\newcommand{\id}{\mathbb{1}}
\newcommand{\Veff}{V_{\rm eff}}
\newcommand{\Vsur}{V_{\rm survey}}
\newcommand{\kmax}{k_{\rm max}}
\newcommand{\kfid}{k_{\rm fid}}
\newcommand{\Nus}{N_{{\nu}_s}}
\newcommand{\cvis}{c^2_{\textrm{vis}}}
\newcommand{\ceff}{c^2_{\textrm{eff}}}
\newcommand{\mnus}{m_{{\nu}_s}}
\newcommand{\ms}{m_{s}^{\textrm{eff}}}
\newcommand{\Omnu}{\Omega_{\nu}h^2}
\newcommand{\mnu}{{\Sigma}m_{\nu}}
\newcommand{\mnuw}{{\Sigma}m_{\nu-w}}
\newcommand{\mnuomk}{{\Sigma}m_{\nu-\Omega_{\rm k}}}

\newcommand{\ns}{\Delta{N}_{\text{eff}}}
\newcommand{\e}[1]{\times 10^{#1}}
\newcommand{\mpcinv}{\, \text{Mpc}^{-1}}

\subsection{Limits on neutrino masses in a non-standard PPS scenario}
\label{sec:neutrpps}
In order to study how the cosmological constraints on the parameters change in more general inflationary scenarios, we assume a non-parametric form for the 
PPS. In particular, we decide to parametrize the scalar PPS with a piecewise cubic Hermite interpolating polinomialâ\cite{Fritsch:1984} (PCHIP) 
%in order 
to avoid some unwanted oscillating behaviour related to the natural cubic spline function
%(for more details, see appendix~A of [\refcite{Gariazzo:2014dla}]).
(see %also 
Appendix~A of Ref. [\refcite{Gariazzo:2014dla}]).
We consider a $\Lambda$CDM model with three degenerate active massive neutrinos together with the PPS approach. We also explore a scenario with three active
light massive neutrinos plus one massive sterile neutrino
species characterized by an effective mass $\ms$. 
%which
%reads: $\ms=(T_s/T_\nu)^3 m_s=(\ns)^{3/4}m_s$,
%being $T_s (T_\nu)$ the current temperature of the sterile (active)
%neutrino species, $\ns=\neff-3.046$
%the effective number of degrees of freedom associated to
%the sterile, and $m_s$ its real mass.

Our baseline data set consists of the
\planck\ 2015 satellite CMB temperature and polarization
APS.\cite{Planck:2015xua,Planck3} %We combine the high-multipole likelihood, $30\leq\ell_{max}\leq 2500,$ with the low-multipole likelihood
%in the range$ \leq2\ell\leq29$  \cite{Planck:2015xua,Planck3}. 
We also consider a prior on the Hubble constant, $H_0$, estimated from a reanalysis of Cepheids data\cite{Efstathiou} and
include measurements of the LSS  in the form of BAO. In particular, we
use the 6dFGS, SDSS-MGS and BOSS DR11 measurements.\cite{Beutler,Ross,Anderson}

\begin{figure}
\includegraphics[width=0.48\textwidth]{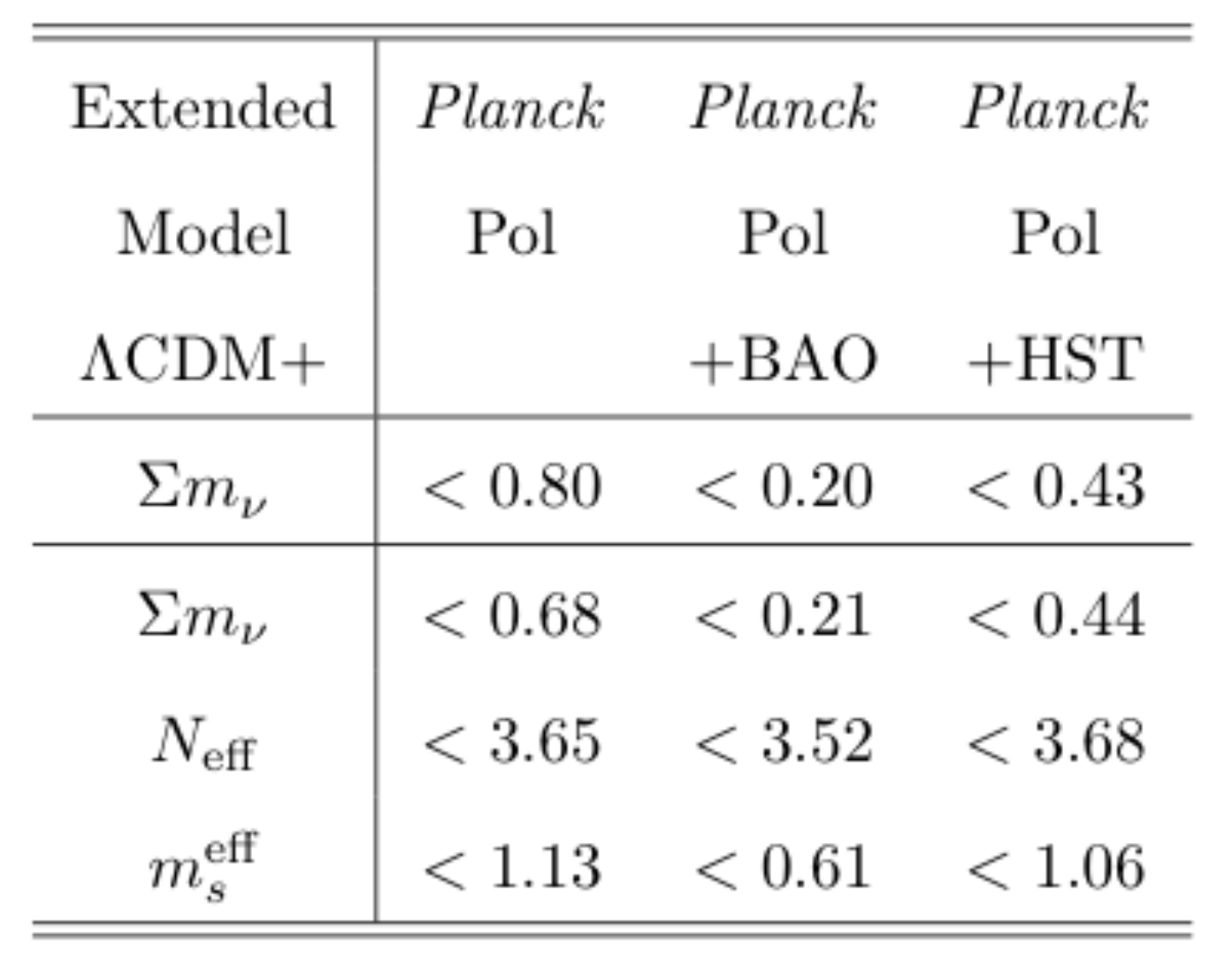}
\includegraphics[width=0.48\textwidth]{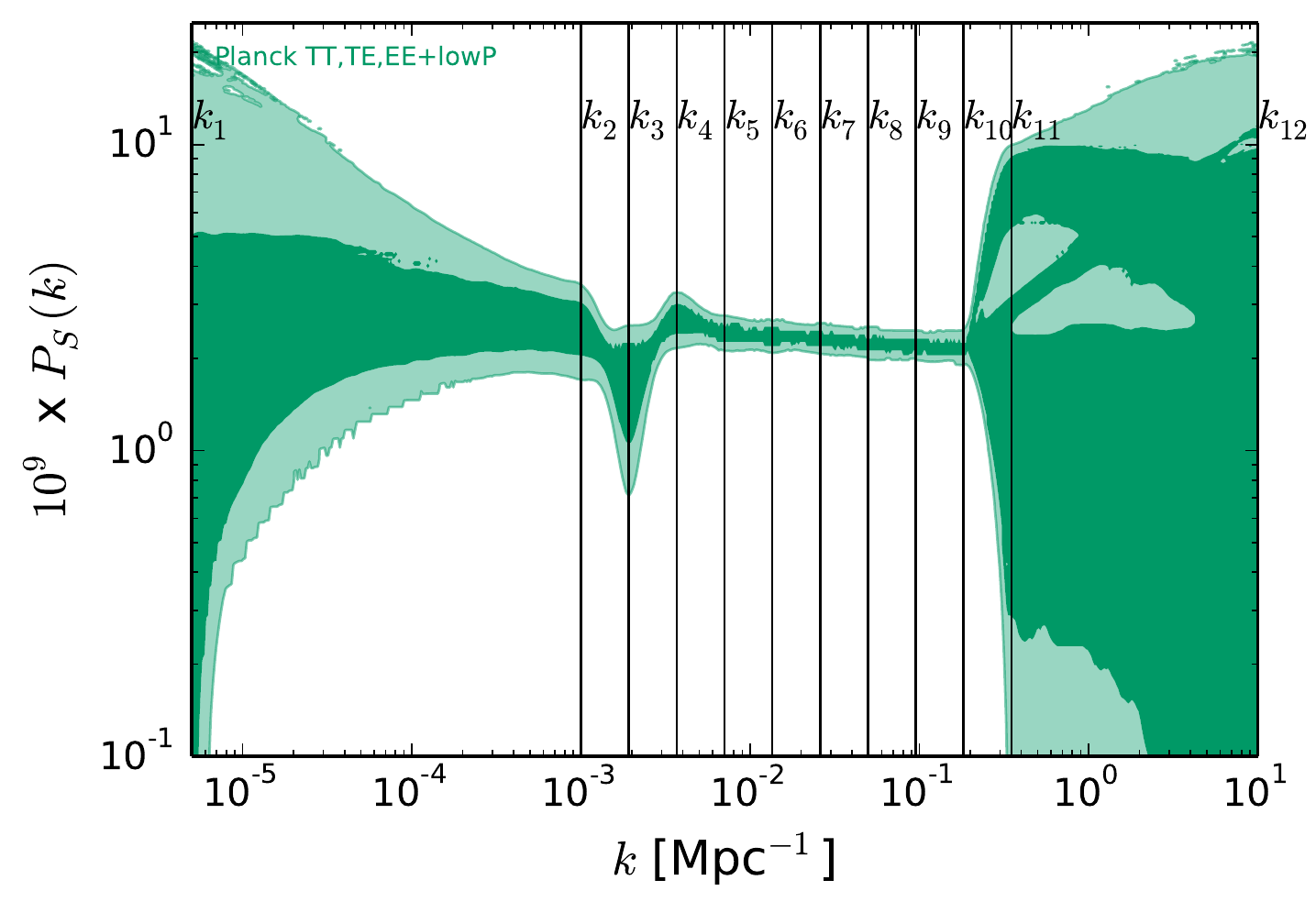}
\caption{Left: 95\% CL on the active (sterile) neutrinos masses and on the total massive neutrino species, $N_\textup{eff}$, from the combination of considered data sets.
Right: $68\%$, $95\%$ and $99\%$~CL allowed regions for the PCHIP PPS scale dependence
in the $\Lambda$CDM$+$${\small {\sum}} m_{\nu}$ model, using CMB data only.}
%in the $\Lambda$CDM+$\mnu$ model, using CMB data only.}
\label{fig:outputPPS}
\vskip -0.3cm
\end{figure}
%\vspace{2.cm}
%%%%%%%%%%%%%%%%%%%%%%%%%%%%%%%%%%%%%%%%%%%%%%%%%%%%%%%%%%

The results are shown in table in the left panel of Fig. \ref{fig:outputPPS}.
%\ref{tab}.
In the first scenario, concerning only CMB measurements, the bound on the sum of massive neutrinos is largely relaxed with respect to the the power-law model
(${\small {\sum}} m_{\nu}<0.49$ eV at 95$\%$ CL).\cite{Planck:2015xua} 
In the second scenario, there is no evidence for neutrino masses nor for non-zero sterile neutrino mass. Concerning only CMB measurements, the bound on the sum of massive neutrinos  is more stringent with respect to previous scenario. The reason for that is due to the degeneracy between ${\small {\sum}} m_{\nu}$ and $\ms$. Notice that in both scenarios the addition of a prior on the Hubble constant and of the BAO data displaces the bounds on ${\small {\sum}} m_{\nu}$ to lower values in agreement with the standard power-law PPS case.\cite{Planck:2015xua} 
%\vspace{0.2mm}
An example of the reconstructed PPS is given in Fig. \ref{fig:outputPPS} (right panel). Note that both $P_{s,1}$ and $P_{s,12}$ are poorly constrained  because of  the absence of measurements at their corresponding wavenumbers.
All the remaining $P_{s,j}$, with $j=2,\ldots,11$ are
well-constrained. In particular, in the range between $k_5$
and $k_{10}$ the PPS can be perfectly described by a power-law parametrization. Moreover we can notice that there is
a significant dip at wavenumbers around $k=0.002\mpcinv$, that comes from the dip at $\ell=20-30$ in the CMB temperature APS and a small bump around $k=0.0035\mpcinv$, corresponding to the increase at $\ell\simeq40$.

\section{Robustness of cosmological thermal axion mass bounds}
\label{axionssec}

Relativistic axions contribute to the dark radiation content of the Universe, increasing the effective number of relativistic degrees of freedom $N_\textup{eff}$ (see Ref. [\refcite{DiValentino:2015zta}] for details), while 
massive thermal axions, when become non-relativistic, affect the LSS formation
suppressing the small scale power, clustering only at scales larger than their free-streaming scale.
Massive axions affect also the CMB temperature anisotropies via the early ISW effect.
All the cosmological axion mass limits\footnote{See e.g. Refs. [\refcite{Giusarma:2014zza,DiValentino:2014zna}] for recent cosmological constraints on thermal and non-thermal axions.} assumed the usual simple power-law description for the primordial perturbations, defined by an amplitude
and a scalar spectral index. 
In Ref. [\refcite{DiValentino:2015zta}] 
the thermal axion mass is constrained using  a non-parametric
description of the scalar perturbation PPS, 
%in order 
to test the robustness of its bounds.
%In particular, 
We adopted a function, the PCHIP\cite{Fritsch:1980} in the same modified version\cite{Gariazzo:2014dla} as in Sect. \ref{sec:neutrpps}, to interpolate the PPS
values in a series of nodes at fixed position.

We discuss here the $\Lambda$CDM model, extended with the axions hot thermal relics, together
with the PPS approach (see Ref. [\refcite{DiValentino:2015zta}] for a similar analysis with two coexisting hot dark matter species, thermal axion and massive neutrinos).
%Our baseline data set consists of 
We consider various CMB measurements: the
temperature data from the \planck\ satellite,\cite{Ade:2013ktc,Planck:2013kta} the
WMAP-9yrs polarization measurements,\cite{Bennett:2012fp} the SPT\cite{Reichardt:2011yv} and ACT\cite{Das:2013zf} datasets. 

Concerning CMB datasets only, the bounds on the thermal axion mass are %totally 
unconstrained in the case in which the PPS is not described by a simple power-law
(see Tab. III in Ref. [\refcite{DiValentino:2015zta}]), while in this last case $m_a<1.83$~eV (see Tab. IV in Ref. [\refcite{DiValentino:2015zta}]).  
Including the Hubble Space Telescope (HST) prior on the Hubble constant,\cite{Efstathiou} $H_0=70.6\pm3.3$ km/s/Mpc,
provides a $95\%$~CL upper limit on the thermal axion mass of $1.31$~eV. 
%Moreover, the 
The further addition of the BAO
measurements\cite{Blake:2011en,Beutler,Percival:2009xn,Padmanabhan:2012hf,Dawson:2012va,Anderson} brings this constraint down to $0.91$~eV, being these last
data sets directly sensitive to the thermal axion free-streaming nature.
%Notice that these 
These %two $95\%$~CL 
upper bounds are very
similar to the ones obtained considering the standard power-law
PPS (see Tab. IV in Ref. [\refcite{DiValentino:2015zta}]). 
%When 
Adding the CFHTLenS\cite{Heymans:2013fya} bounds on the $\sigma_8$-$\Omega_m$ relationship, the thermal axion mass bounds become weaker $m_a<1.29$~eV, since this dataset prefers a lower $\sigma_8$ value. 
Finally, %when 
considering the \planck\ Sunyaev-Zeldovich (PSZ) 2013 catalogue\cite{Ade:2013lmv} dataset with fixed cluster mass bias, $\sigma_8 (\Omega_m/0.27)^{0.3}=0.78\pm 0.01$, together with the CMB, BAO and HST measurements, a non-zero thermal axion mass of $\sim 1$~eV is
favored at $\sim4\sigma$ level. Using more realistic approaches for the cluster mass bias,\cite{Ade:2013lmv} $\sigma_8 (\Omega_m/0.27)^{0.3}=0.764\pm 0.025$, the errors on the so-called cluster normalization condition are larger, and, consequently, the preference for a non-zero axion mass is reduced.

\begin{figure}
\begin{tabular}{c c}
\includegraphics[width=5cm]{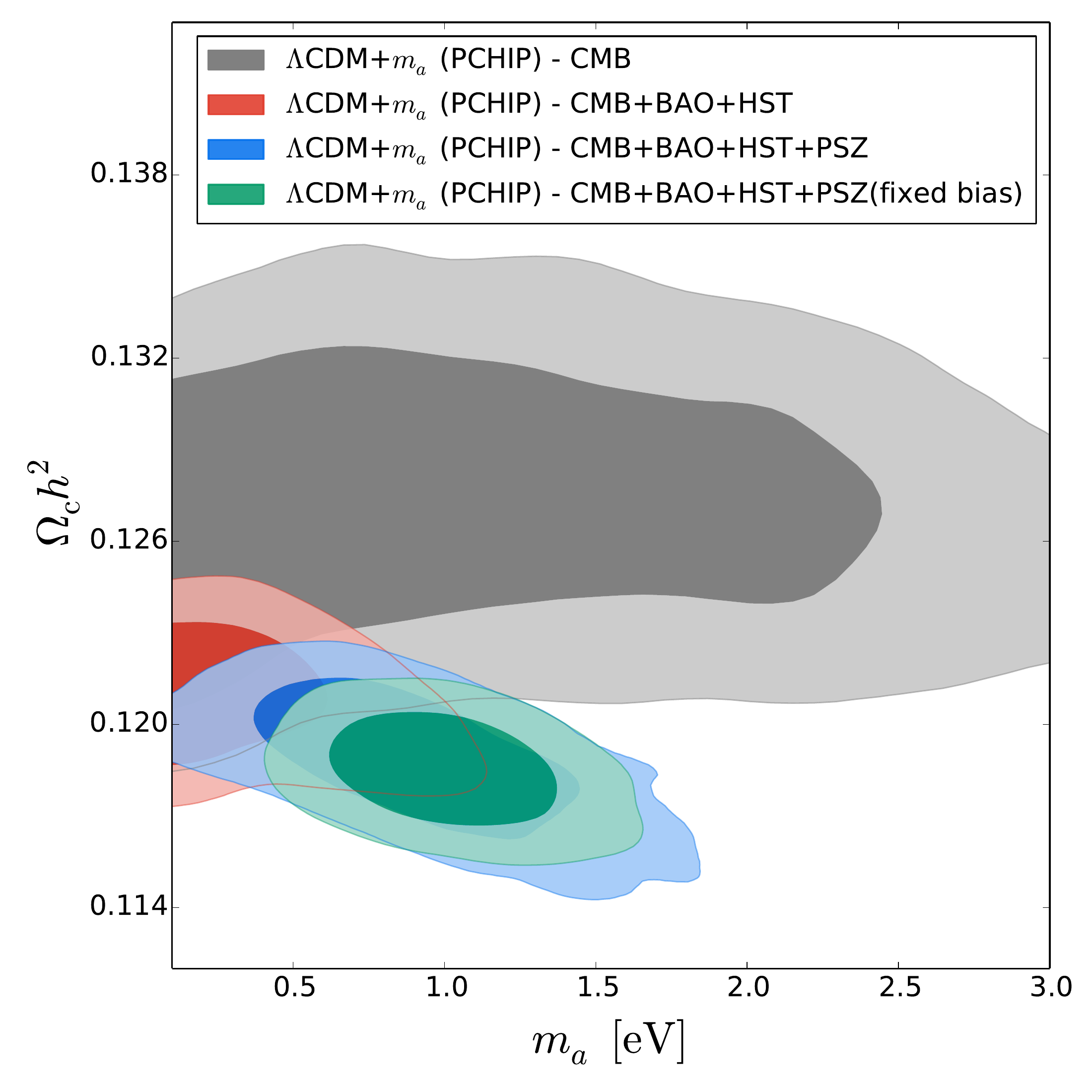}&\includegraphics[width=5cm]{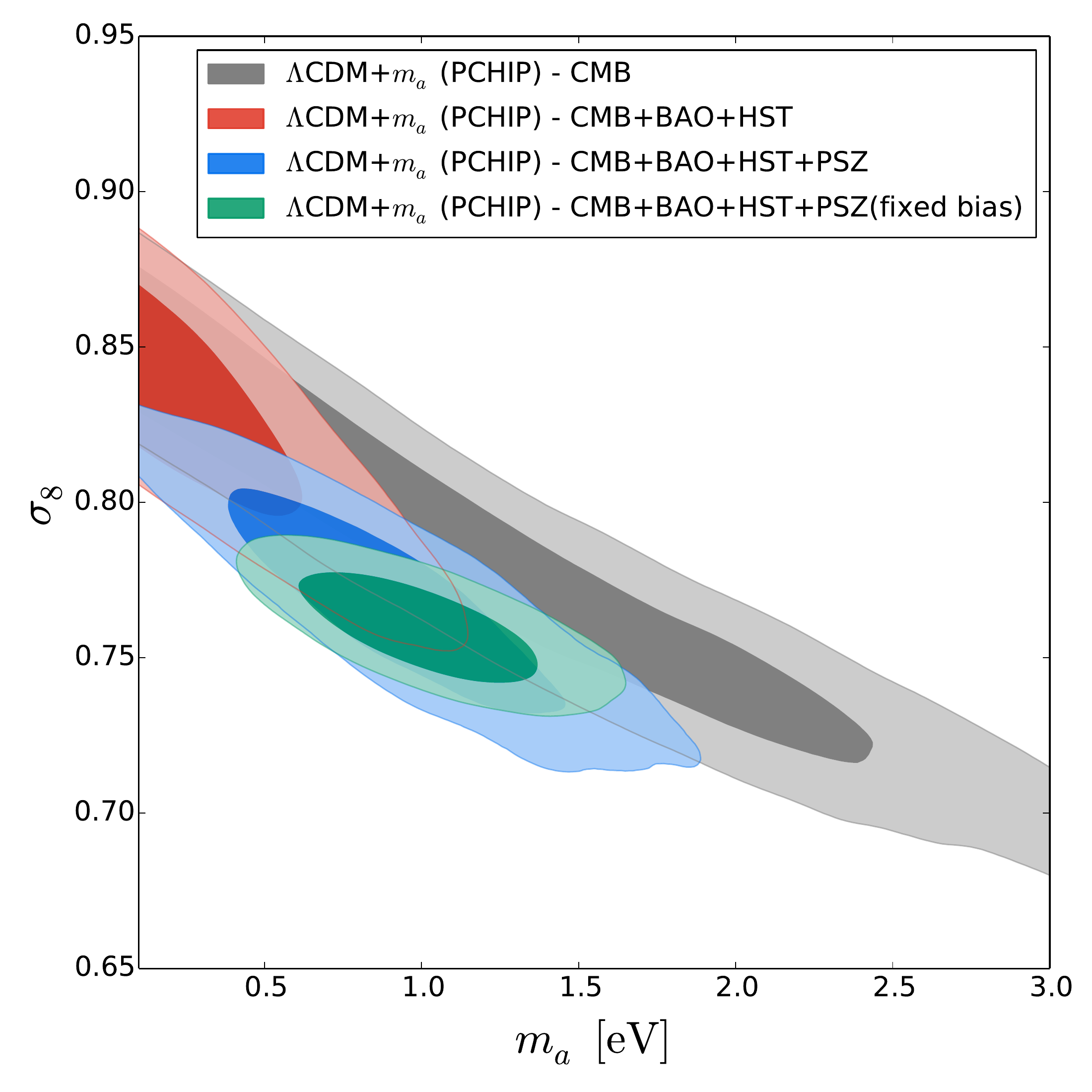}\\
\end{tabular}
\vskip -0.2cm
 \caption{$68\%$ and $95\%$~CL allowed
   regions in the ($m_a$, $\Omega_c h^2$) plane (left panel) and in the
   ($m_a$, $\sigma_8$) plane (right panel) for different 
   %possible
   data combinations, when a PCHIP PPS is assumed. From Ref. [94].}
%   From [\refcite{DiValentino:2015zta}].}
\label{fig:ma_pchip}
\vskip -0.3cm
%\vskip -0.7cm
\end{figure}

Our results are summarized in Fig. \ref{fig:ma_pchip}. In conclusion, using a 
%the PCHIP 
non-parametric description\cite{Gariazzo:2014dla} of the scalar perturbation PPS that relaxes
the power-law assumption in Ref. [\refcite{DiValentino:2015zta}], we tested the robustness of the cosmological axion mass
bounds, found to be only mildly sensitive to the PPS
choice and therefore not strongly dependent on the
particular details of the underlying inflationary model. 

%\begin{figure*}[!t]

\section{Cosmological constraints on the neutron lifetime}
\label{neutronsec}

The study of the neutron lifetime, $\tau_n$, a fundamental quantity in nuclear physics, is fascinating since the current status of particle physics experiments is still puzzling and unclear.
The current used value is the one quoted by the
Particle Data Group\cite{PDG} $\tau_n=(880.3 \pm 1.1) \, \text{s}$ and it is obtained as an average between the seven most recent experiments, bottle-method and beam-method like
(for further details see [\refcite{Pignol}]). Combining the five most recent bottle-method measurements one obtain the tight constraint $\tau_n= (879.6 \pm 0.8) \, \text{s}$ while from the two most recent beam-method measurements one obtain $\tau_n= (888.0 \pm 2.1) \, \text{s}$. Given this tension, it is 
interesting to investigate if 
cosmological measurements can constrain the neutron lifetime in an independent way with respect to particle physics experiments, % and 
thus testing them,
and, moreover, to address the implications for cosmology of a 
a more precise determination of the neutron lifetime.

We start discussing constraints on $\tau_n$ from current cosmological data. Assuming Standard Big Bang Nucleosynthesis it is possible to evaluate primordial abundances of light elements from CMB as functions of few parameters:\cite{parthenope} the baryonic abundance, the relativistic degrees of freedom, the chemical potential of electron neutrinos and the neutron lifetime. 
Neglecting the chemical potential, considering the high precision achieved in the determination of baryonic abundance and fixing 
$N_{\text{eff}}$ 
%the relativistic degrees of freedom 
to its standard value of $3.046$, from primordial abundances (in particular Helium abundance) we can infer the value of the neutron lifetime. 
We start analyzing \planck\ 2015 results as CMB dataset 
with the publicly available MCMC package \texttt{cosmoMC}. 
Table \ref{table:CMB-taun-fore} reports the most interesting results (for complete analysis see Ref. [\refcite{Salvati:2015wxa}]).

The next step is to combine CMB observations with direct astrophysical measurements of Helium. %Moreover, 
We consider eight primordial Helium measurements collected in the last ten years and combine them with \planck\ data and select two possible independent combinations of these astrophysical datasets 
(referred to as M12-P for Refs. [\refcite{2014ApJ...786...14M,Peimbert:2007vm}] 
and M12-I14 for Refs. [\refcite{2014ApJ...786...14M,Izotov:2014fga}]).
As shown in Table \ref{table:CMB-taun-fore}, combining the constraining power of CMB data, sensitive to the baryon density, with the Helium astrophysical measurements we obtain more stringent limits on the neutron lifetime, with respect to cosmological data only. 

We extend the analysis performing some forecasts on future cosmological experiments. Considering that CMB sensitivity on $\tau_n$ is encoded in the small-scale region,
we expect tighter constraints from next CMB projects planned to measure the high $\ell$ range. 
As reported in Table \ref{table:CMB-taun-fore}, the most stringent constraint is obtained by the combination of future experiments COrE\footnote{http://www.core-mission.org/}
and {\it Euclid}, giving $\tau_n= (880.3 \pm 6.7) \, \text{s}$.

\begin{table}
%\verb|\tbl{Values of $\tau_n$ at 68 \% c.l. for the different cosmological and astrophysical datasets (see text).}|
%\verb|{\begin{tabnote}Values of $\tau_n$ at 68 \% c.l. for the different cosmological and astrophysical datasets (see text).\end{tabnote}}|
\tbl{Values of $\tau_n$ with 1$\sigma$ erros for cosmological and astrophysical datasets.}
%\begin{center}
%\scalebox{0.8}{
{\begin{tabular}{|c|c|}
\hline
Dataset & $ \bf \tau _n \, [\text{s}]$ \\
\hline
\hline
\planck\ + BAO + Lensing & $894 \pm 63$\\
\hline
\hline
M12-I14 & $905.7 \pm 7.8$\\
M12-P & $886.7 \pm 8.8$ \\
\hline
\hline
COrE & $ 880 \pm 11$\\
CVL & $880.7 \pm 5.5$\\
COrE + {\it Euclid} & $880.3 \pm 6.7$ \\
\hline
\end{tabular}%}
}
%\caption{\footnotesize{Values of $\tau_n$ at 68 \% c.l. for the different cosmological and astrophysical datasets (see text).}}
\label{table:CMB-taun-fore}
\vskip -0.3cm
%\end{center}
\end{table}

In conclusion, the combination of CMB anisotropies and astrophysical observations allows to obtain stringent limits and shed light on the present experimental discrepancies, 
while future cosmological missions, such as COrE and {\it Euclid}, could reach a sensitivity comparable with that of current experiments. %al uncertainties. 

\section{Testing general relativity with cosmic polarization rotation}
\label{cprsec}

The CPR provides a test of the EEP, 
which is the foundation of any metric theory of gravity, including GR.
Almost all the information 
about the Universe outside the solar system is carried to us by photons,
with their direction, 
energy and polarization. The latter consists essentially in the position angle (PA) of the polarization ellipse, i.e. photons carry throughout the Universe an important geometrical information. 
To properly use this information,
it is important to know if and how it is changed while photons travel to us. 
The directions of photons can be modified by gravitational fields
and their energies are modified by the Universe expansion, while
the polarization PA is 
modified while photons travel in a plasma with a magnetic field, the so called Faraday rotation, 
proportional to the wavelength squared.
Is the polarization PA also modified while photons travel large distances {\it in vacuum}?
Searches for CPR deal with this important question.

Clearly, 
if the CPR angle $\alpha$ is not zero, it should be either positive for a counter-clockwise rotation, or negative for a clockwise rotation (we adopt the IAU convention\cite{iau74} for PA, which increases counter-clockwise facing the source, from North through East). Therefore symmetry must be broken at some level, leading to the violation of fundamental physical principles (see Ref. [\refcite{Nio10}] for a recent review).
Indeed CPR is linked also to a possible violation of the EEP.
The reasons for the link are due to the unique counterexample to Schiff's conjecture\cite{Sch60} that any consistent Lorentz-invariant theory of gravity which obeys the weak equivalence principle (WEP) would also obey the EEP, which involves a pseudoscalar field, producing CPR.\cite{Nio77} Therefore, if we could show that 
$\alpha=0^{\circ}$, 
the EEP would be tested with the same high accuracy of the WEP, greatly increasing our confidence in the EEP and then in GR.
See Ref. [\refcite{diS15}] for a recent review of CPR tests.

\begin{figure}
\begin{center}
\includegraphics[width=5.5cm]{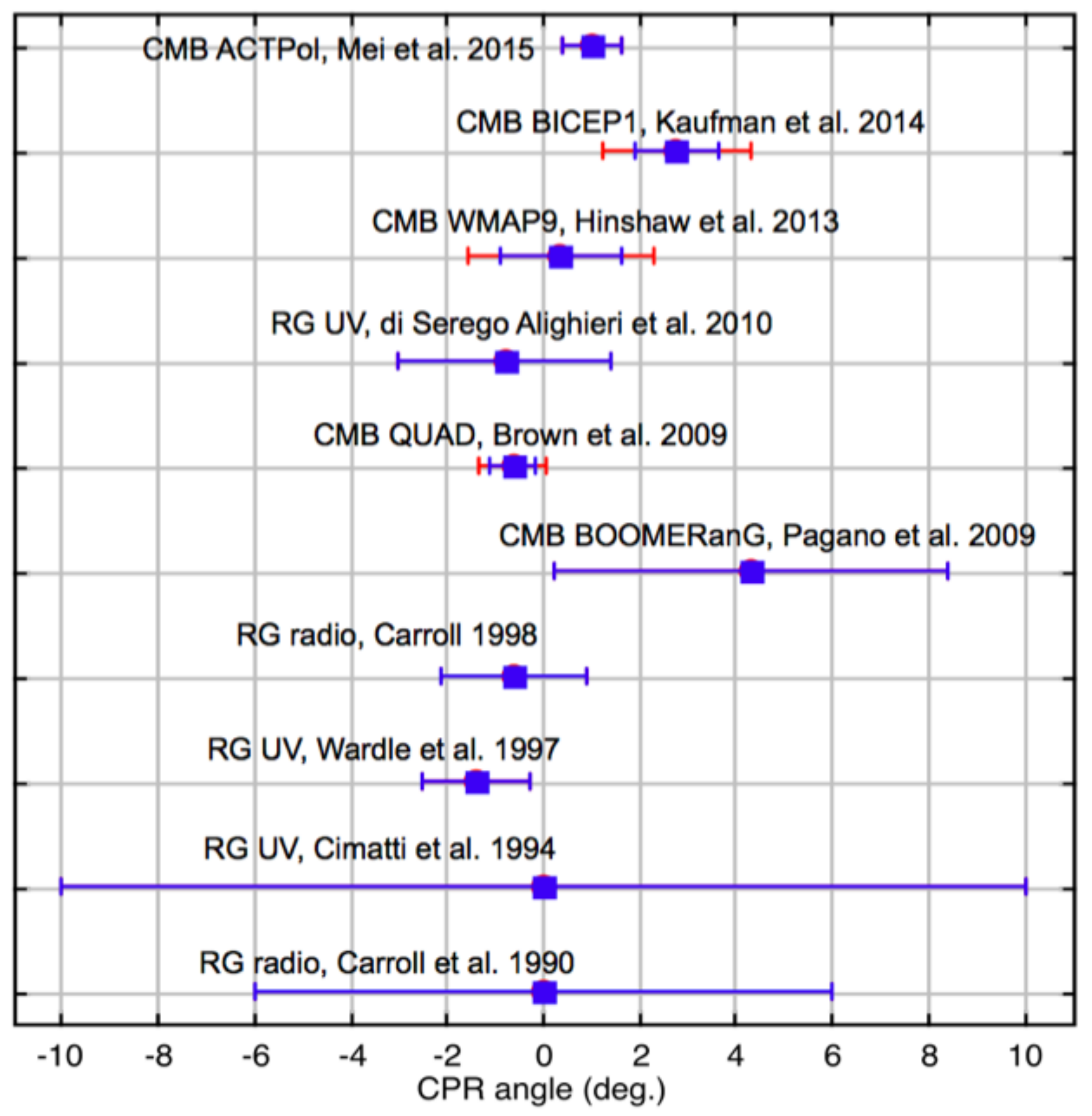}
\end{center}
\vskip -0.2cm
\caption{CPR angle measurements by the various experiments, displayed in chronological order. Blue error bars: statistical errors; red error bars: including also systematics, if present/available. 
A systematic error should be added to the ATCPol measurement, equal to the unknown difference of the Crab Nebula polarization PA between 146 GHz and 89 GHz.}
\label{aba:fig1}
\vskip -0.3cm
\end{figure}

CPR tests are simple in principle: they require a distant source of polarized radiation 
with established polarization orientation at the emission, $PA_{em}$. 
By measuring the observed orientation $PA_{obs}$, the CPR angle can be calculated:
\begin{equation}
%$$
\alpha  = PA_{obs} - PA_{em}. 
%$$
\end{equation}
The problem is the estimate of $PA_{em}$. Fortunately it can be solved using the fact that scattered radiation is polarized perpendicularly to the plane containing the incident and scattered rays. This simple physical law has been applied to CPR tests, using both the ultraviolet (UV) radiation of radio galaxies (RG) and the tiny anisotropies of the CMB. The first CPR tests 25 years ago used instead a statistical analysis of the radio
polarization in RG.\cite{Car90} The most accurate CPR tests obtained with the various methods are summarized in Fig.~\ref{aba:fig1}, based on data in Ref. [\refcite{diS15}].

In summary, the results so far are consistent with a null CPR with upper limits of the order of one degree.

%\subsection{Current problems in searches for Cosmic Polarization Rotation}
\subsection{Current problems and future prospects}

Searches for CPR using the UV polarization of RG have reached the limits allowed by current instrumentation,
for the lack of RG suitable for the test and bright enough so that their polarization can be measured with the available instruments.

The most accurate results are now obtained with the CMB polarization, averaging over large sky areas, i.e. assuming uniform CPR over these areas. A current problem with CPR searches using the CMB is the calibration of the polarization PA for the lack of sources with precisely known PA at CMB frequencies. This introduces a systematic error, 
similar to the statistical measurement error, of 
about $1^{\circ}$
(see Fig.~\ref{aba:fig1}). 
Recently, the polarization PA of the Crab Nebula (Tau $\alpha$) has been measured with an accuracy of $0.2^{\circ}$ at 89.2 GHz.\cite{aumont} However, most CMB polarization measurements are made at 100-150 GHz
and the Crab Nebula is not visible from the South Pole, the site of several CMB experiments. In order to overcome the PA calibration problem, some CMB polarization experiments have used a $TB$ and $EB$ nulling procedure,\cite{Kea13} but 
this would eliminate together the PA systematic error and any CPR angle $\alpha$, so it cannot be used for CPR tests.

Furthermore, we note that,
unfortunately, 
in the 
pixelization tool\cite{Gor05} widely adopted in CMB experiments\footnote{See Ref. [\refcite{Wal15}] for a pixelization software adopting IAU convention.} 
the polarization PA is assumed to increase clockwisely (looking at the source), which is opposite to the standard IAU convention,
adopted in other bands,
thus 
calling for caution when comparing measures with different methods, like for CPR tests.

The different methods are complementary in many ways. They cover different wavelength ranges and the methods at shorter wavelength have an advantage, if CPR effects grow with photon energy,
as foreseen in some cases.\cite{Kos01, Kos02} They also reach different distances, and the CMB method obviously reaches furthest. However the relative difference in light travel time between z = 3 and z = 1100 is only 16\%.

Improvements are
expected by better targeted high resolution radio polarization measurements of RGs and quasars, by more accurate UV polarization measurements of RGs with the coming generation of giant optical
telescopes,\cite{deZ14, San13, Ber14} and by future CMB polarization measurements such as those from {\it Planck}\cite{Ade:2013zuv} and BICEP3.\cite{Ahm14} 
Indeed, the \planck\ satellite 
has
a very low statistical error ($\sim 0.06^{\circ}$) for CPR measurements, 
but to exploit its great accuracy a significant reduction in 
the systematic error in the calibration of the polarization angle (currently of $\sim 1^{\circ}$ for CMB polarization experiments) is needed (see also Ref. [\refcite{gruppuso_biref}]).
Great opportunities will 
come from 
%be offered by 
more precise polarization measurements 
%of the polarization angle 
of celestial sources at CMB frequencies with ATCA\cite{Mas13} and ALMA,\cite{Tes13} 
and by a calibration source on a satellite.\cite{Kau15}

\def\lsim{\,\lower2truept\hbox{${< \atop\hbox{\raise4truept\hbox{$\sim$}}}$}\,}
\def\gsim{\,\lower2truept\hbox{${> \atop\hbox{\raise4truept\hbox{$\sim$}}}$}\,}

%%%%%%%%%%%%%%%%%% now a standard article style for the most part

\section{SKA contribution to future CMB spectrum experiments}
\label{sec:cmb_spect}

\begin{figure}
\minipage{0.45\textwidth}
\vskip -1.cm
%\hskip -0.7cm
\includegraphics[width=\linewidth]{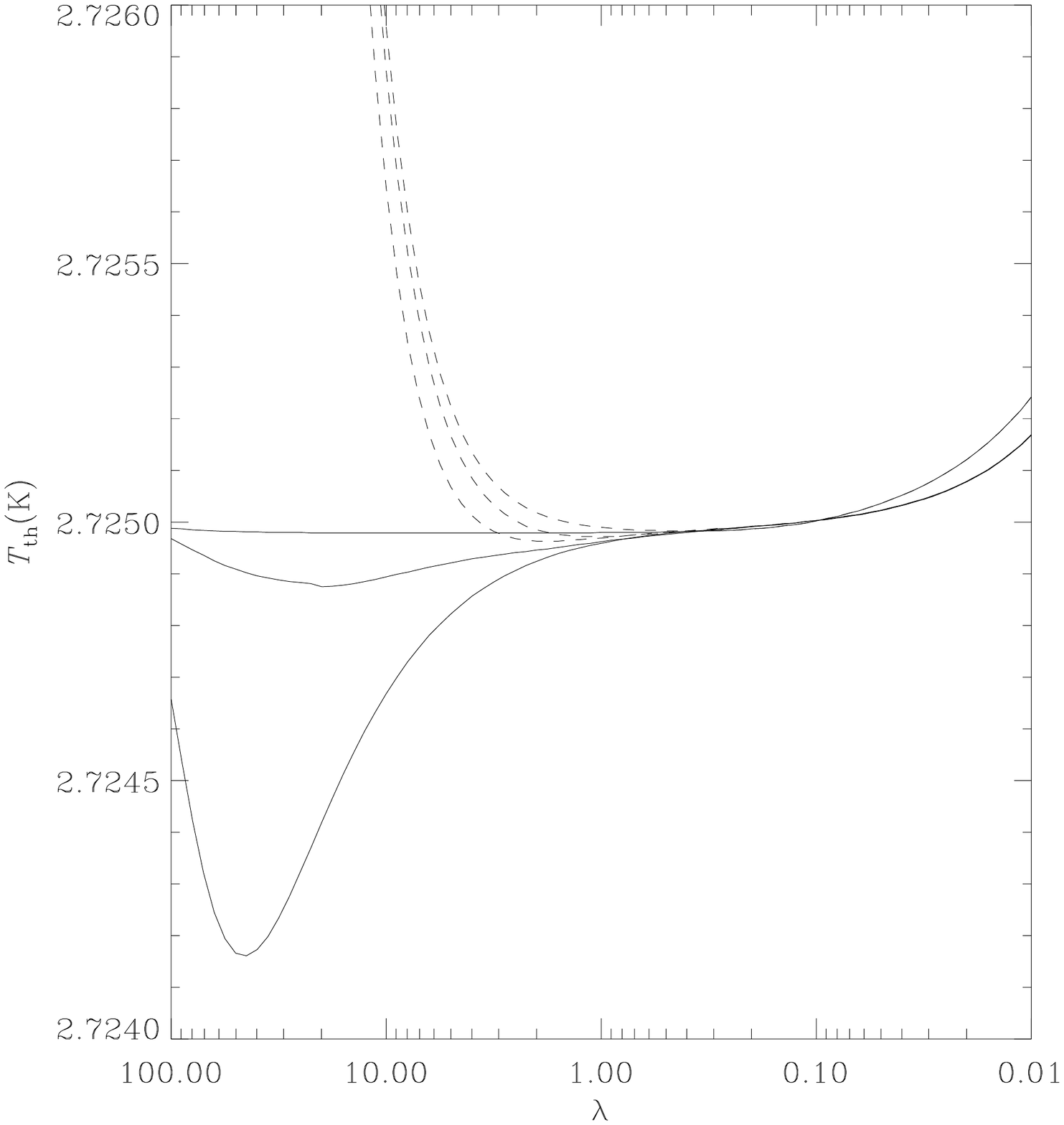}
\endminipage\hfill
\minipage{0.45\textwidth}
\vskip -1.cm
\hskip -0.7cm
\includegraphics[width=\linewidth]{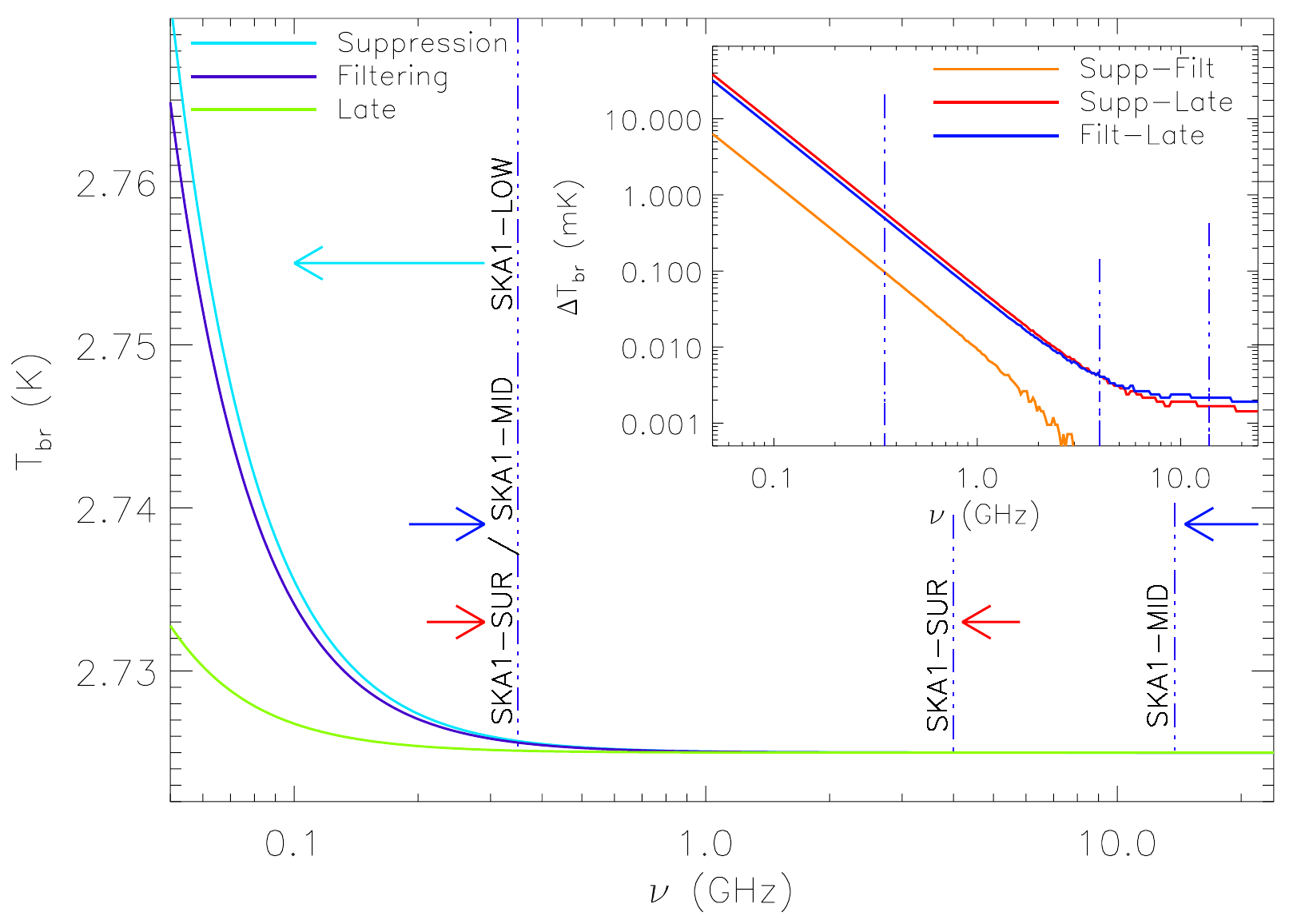}
\endminipage\hfill
\vskip -1.cm
\caption{Left panel: distorted spectra 
in equivalent thermodynamic temperature
vs. 
$\lambda$ (cm) with
late energy injection 
$\Delta \varepsilon / \varepsilon_i = 5 \times 10^{-6}$
plus an early/intermediate
energy injection $\Delta \varepsilon / \varepsilon_i = 5 \times 10^{-6}$
($\sim$ 20 times smaller than current upper limits) at the ``time'' Comptonization 
parameter $y_h=5, 1, 0.01$ (bottom to top;
the cases at $y_h=5$
and $1$ are very similar at short $\lambda$; solid lines) 
plus a FF distortion
with $y_B=10^{-6}$ (dashes).
$y_h = y$
with $T_e=T_{CMB}$ when the integral
is computed from the energy injection time
to the current time. 
Right panel: FF distortion 
in SKA2 frequency range
by two astrophysical reionization histories 
(a {\it late} phenomenological model
is also displayed for comparison).
Inset: models absolute differences; vertical lines: ranges of SKA1 configurations. From Ref. [141].}
\label{fig:cmb_dist}
\vskip -0.3cm
\end{figure}

Recent limits on CMB spectral distortions and constraints on
energy dissipation processes in the plasma\cite{SB02}
are mainly set by
COBE/FIRAS experiment.\cite{mather90} 
High accuracy CMB spectrum space experiments, such DIMES
($\lambda \gsim 1$~cm) and
FIRAS~II ($\lambda \lsim 1$~cm), were proposed to
constrain energy exchanges up to 100 times better than FIRAS.
Dissipation processes at early times ($z \gsim 10^5$) result in 
Bose-Einstein (BE)-like distortions,\cite{SZ70} while late epochs mechanisms ($z \lsim 10^4$) before or after the
recombination era generate Comptonization and free-free (FF)
distortions.\cite{bart_stebb_1991} 
New space missions 
were proposed to probe cosmic origin and evolution observing CMB temperature and polarization anisotropies 
with $\sim$ degree resolution, as in PIXIE\cite{kogutpixie} and LiteBird,\footnote{http://litebird.jp/eng/} 
or with arcmin resolution, as in COrE,
PRISM\footnote{http://www.prism-mission.org/} and COrE+, in combination with  
spectrum measurements in the case of PIXIE and PRISM. 
SKA extremely high sensitivity and resolution can contribute to set new constraints on CMB spectral distortions
beyond current limits. Improved absolute temperature measures
will strengthen the constraints on CMB spectrum affected by (pre-recombination) decaying and annihilating particles, by superconducting cosmic strings
electromagnetic radiation, by energy injection of evaporating primordial black holes (BH). Spectral distortions could constrain non evaporating BH spin,
small scale magnetic fields, vacuum energy density decay, axions. In general, departure of CMB spectrum from a perfect blackbody is
theoretically predicted by:\cite{sunyaevkhatri2013} ({\it i}) cosmological reionization, producing electron heating
and physically correlated Comptonization distortion (with typical Comptonization parameter $y \simeq (1/4) \Delta \varepsilon / \varepsilon_i  \approx 10^{-7}-10^{-6}$
proportional to the fractional energy exchanged in the interaction), 
and free-free (FF) distortion;
({\it ii}) dissipation of primordial perturbations at small scales,
damped by photon diffusion and thus invisible in CMB anisotropies,
produces BE-like distorted spectra characterized by a positive chemical potential 
$\mu_{0} \simeq 1.4 \Delta\varepsilon/\varepsilon_{i} \approx 10^{-9} - 10^{-7} $; ({\it iii}) BE condensation of CMB photons by colder electrons associated with the matter temperature decrease in the expanding Universe 
relatively faster than that of radiation gives $\mu_{0}  \approx -3 \times 10^{-9}$. The above FF signal 
is the most relevant type of low-frequency global spectral distortion (see Fig. \ref{fig:cmb_dist}). Indeed, the FF term is proportional to the square of baryon density and the structure formation process implies a rate amplification 
 by a factor $\simeq 1 + \sigma^{2}$ (being $\sigma^{2}$ the matter distribution variance) with respect to the case of homogeneous plasma.\cite{TrombettiBurigana2014} 
SKA high sensitivity and resolution can also be used to model the contribution from 
Galactic emissions and extragalactic foreground, a fundamental step to accurately observe these kinds of distortions. 
Extragalactic source contribution is small compared to Galactic 
radio emission, currently the major astrophysical problem in CMB spectrum experiments,
but, unlike the Galactic emission, it cannot be subtracted from the CMB monopole 
temperature by exploiting its angular correlation properties. 
A direct radio background estimate from precise number counts will certainly improve the robustness
of this kind of analyses. Exploiting the recent differential number counts
 at 0.153 GHz,
 %\cite{williamsetal2013} 
 0.325 GHz,
 %\cite{mauchetal2013}
1.4 GHz,%\cite{condonetal2012}
and 1.75 GHz % \cite{vernstrometal2014} 
it is possible to evaluate the contribution, $T_{b}$, to the radio background from extragalactic sources in various
ranges of flux densities. 
These signals can be significant at the accuracy level 
potentially achievable with future experiments. 
Subtracting sources brighter than several tens of nJy,
$T_{b} \lsim 1$\,mK at $\nu \gsim 1$\,GHz, but $T_{b} \gsim 10$\,mK below $0.3$\,GHz.
The minimum source detection threshold is given by  
the source confusion noise. 
The finite angular extension of faint galaxies, $\theta \sim 1''$, implies a ``natural confusion limit`` $ \sim 10$\,nJy at $\nu \sim 1.4$\,GHz, not a relevant limitation for deep surveys.\cite{condonetal2012} 
At $1\, {\rm GHz} \, \lsim \nu \lsim {\rm some \, \, GHz}$
($\lambda \approx 1$\,dm) the signal amplitudes found for CMB distorted spectra well below FIRAS constraints 
%(see Fig.~\ref{fig:cmb_dist}) 
are significantly larger than the estimates of the background from extragalactic 
sources fainter than some tens of nJy. At decreasing frequencies FF distortion amplitude increases but, at the same time, source confusion noise may represent a serious problem, 
possibly preventing the achievement of the faint detection threshold necessary to have a source contribution to the background significantly 
less than the CMB distortion amplitude.
%%%%%%%%%%%%%%%%%%

%%%%%%%%%%%%%%%%%%
%\begin{figure}[t]
%%%%%%%%%%%%%%%%%%

SKA will trace the neutral hydrogen distribution and the neutral-to-ionized transition state
at the reionization epoch through the 21-cm line. %\cite{Schneideretal2008}. 
It could trace the development of ionized material directly by looking for FF emission from ionized halos.  
The expected signal can be derived by reionization models through both semi-analytical
methods\cite{naselskychiang04} and numerical simulations.\cite{ponenteetal2011} 
Dedicated high resolution sky areas observations allow to distinguish FF distortion by ionized halos rather than by diffuse ionized IGM.
SKA should be able to detect up to $\sim 10^{4}$ individual FF emission sources with $z>5$ in 1 deg$^2$
%$\Box$,
discerning ionized halos or diffuse ionized IGM FF distortions.
Thus, the precise mapping of individual halos represents an interesting goal for the excellent imaging capabilities of SKA.

\bigskip

{\small
\noindent
{\bf Acknowledgements --}  
The authors that are members {\it Planck} Collaboration warmly thank the {\it Planck} Collaboration
for numberless and constructive conversations on the subjects
discussed here.
Some of the results in this paper have been derived using the HEALPix\cite{Gor05} package.
We acknowledge the use of the NASA Legacy Archive for Microwave Background Data Analysis (LAMBDA) 
and of the ESA {\it Planck} Legacy Archive.
CB, AG, ML and TT acknowledge partial support by ASI/INAF Agreement 2014-024-R.0 for the
{\it Planck} LFI Activity of Phase E2.
}

\end{document}